\documentclass[sigplan]{acmart}

\setcopyright{none}
\renewcommand\footnotetextcopyrightpermission[1]{} 
\AtBeginDocument{%
  }

\usepackage{amssymb}
\usepackage{times}
\usepackage{amsfonts}

\usepackage{subfig}
\usepackage{wrapfig}
\usepackage{float}
\usepackage{multirow, graphicx}
\usepackage{algorithm}
\usepackage{algpseudocode} %algorithm: use algorithmicx package, to refer to http://en.wikibooks.org/wiki/LaTeX/Algorithms_and_Pseudocode#Typesetting_using_the_algorithmicx_package
\usepackage{amsmath}

\usepackage{xspace}

\usepackage{url}

\usepackage{breakurl}

\usepackage{sty/widetext} %show equation in two columns

\usepackage{stackrel}

\ifdefined\TTSTYLETARGETreport
\usepackage[
    backend=bibtex,
    maxbibnames=100,
    firstinits=true,
    bibstyle=numeric,
    citestyle=numeric
]{biblatex}
\fi

\usepackage{pifont}% http://ctan.org/pkg/pifont
\newcommand{\cmark}{\ding{51}}%
\newcommand{\xmark}{\ding{55}}%

\usepackage{todonotes}

\newcommand{\ignore}[1]{}

\usepackage{graphicx, multirow}

\usepackage{tabularx, multirow, rotating} %tables, dependent on multirow package.
\usepackage{subfig} %figures
\usepackage{algorithm} %algorithm
\usepackage{algpseudocode} %algorithm: use algorithmicx package, to refer to http://en.wikibooks.org/wiki/LaTeX/Algorithms_and_Pseudocode#Typesetting_using_the_algorithmicx_package
\usepackage{ulem} % strike out text
\usepackage{color} % colored text

\usepackage{listings} 
\lstdefinestyle{Oracle}{basicstyle=\ttfamily,
                        keywordstyle=\lstuppercase,
                        emphstyle=\itshape,
                        showstringspaces=true,
                        }
\makeatletter
\newcommand{\lstuppercase}{\uppercase\expandafter{\expandafter\lst@token
                           \expandafter{\the\lst@token}}}
\newcommand{\lstlowercase}{\lowercase\expandafter{\expandafter\lst@token
                           \expandafter{\the\lst@token}}}
\makeatother

\usepackage{tikz}

\newif\ifboldnumber

\algrenewcommand\alglinenumber[1]{%
  \footnotesize\ifboldnumber\bfseries\fi\global\boldnumberfalse#1:}

\usepackage{color, colortbl}
\definecolor{Gray}{gray}{0.9}
\definecolor{LightCyan}{rgb}{0.88,1,1}
\usepackage[first=0,last=9]{lcg}

\usepackage{colortbl}
\usetikzlibrary{calc}
\usepackage{zref-savepos}

\newcounter{NoTableEntry}
\renewcommand*{\theNoTableEntry}{NTE-\the\value{NoTableEntry}}

\usepackage{makecell}

\begin{document}

\title{Efficiently Hardening SGX Enclaves against Memory Access Pattern Attacks via Dynamic Program Partitioning}

%% used to denote shared contribution to the research

\author{Yuzhe Tang}
\affiliation{%
  \institution{Syracuse University}
  \city{Syracuse}
  \state{New York}
  \country{USA}
}
\email{ytang100@syr.edu}

\author{Kai Li}
\affiliation{%
 \institution{San Diego State University}
 \city{San Diego}
 \state{California}
 \country{USA}
}
\email{kli5@sdsu.edu}

\author{Yibo Wang}
\affiliation{%
 \institution{Syracuse University}
 \city{Syracuse}
 \state{New York}
 \country{USA}
}
\email{ywang349@syr.edu}

\author{Jiaqi Chen}
\affiliation{%
 \institution{Syracuse University}
 \city{Syracuse}
 \state{New York}
 \country{USA}
}
\email{jchen217@syr.edu}

\author{Cheng Xu}
\affiliation{%
 \institution{Hong Kong Baptist University}
 \city{Hong Kong}
}
\email{chengxu@comp.hkbu.edu.hk}

%% If needed use a foot or author note to identify equal contribution

\begin{abstract}
  Intel SGX is known to be vulnerable to a class of practical attacks exploiting memory access pattern side-channels, notably page-fault attacks and cache timing attacks. A promising hardening scheme is to wrap applications in hardware transactions, enabled by Intel TSX, that return control to the software upon unexpected cache misses and interruptions so that the existing side-channel attacks exploiting these micro-architectural events can be detected and mitigated. However, existing hardening schemes scale only to small-data computation, with a typical working set smaller than one or few times (e.g., $8$ times) of a CPU data cache.

This work tackles the data scalability and performance efficiency of security hardening schemes of Intel SGX enclaves against memory-access pattern side channels. The key insight is that the size of TSX transactions in the target computation is critical, both performance- and security-wise. Unlike the existing designs, this work {\it dynamically} partitions target computations to enlarge transactions while avoiding aborts, leading to lower performance overhead and improved side-channel security. We materialize the dynamic partitioning scheme and build a C++ library to monitor and model cache utilization at runtime. We further build a data analytical system using the library and implement various external oblivious algorithms. Performance evaluation shows that our work can effectively increase transaction size and reduce the execution time by up to two orders of magnitude compared with the state-of-the-art solutions.

\end{abstract}

\maketitle
\pagestyle{plain}
\newcommand{\yz}[1]{\footnote{\textcolor{red}{(Yuzhe: #1)}}}

\providecommand{\eoa}{\textsf{EOAlgo}\xspace}

\definecolor{mygreen}{rgb}{0,0.6,0}
\definecolor{mymauve}{rgb}{0.88,0.69,0}
\lstset{ %
  backgroundcolor=\color{white},   % choose the background color; you must add \usepackage{color} or \usepackage{xcolor}
  %basicstyle=\footnotesize\ttfamily,        % the size of the fonts that are used for the code
  basicstyle=\scriptsize\ttfamily,        % the size of the fonts that are used for the code
  breakatwhitespace=false,         % sets if automatic breaks should only happen at whitespace
  breaklines=true,                 % sets automatic line breaking
  captionpos=b,                    % sets the caption-position to bottom
  commentstyle=\color{mygreen},    % comment style
  deletekeywords={...},            % if you want to delete keywords from the given language
  escapeinside={\%*"}{"*)},          % if you want to add LaTeX within your code
  extendedchars=true,              % lets you use non-ASCII characters; for 8-bits encodings only, does not work with UTF-8
  %frame=single,                    % adds a frame around the code
  keepspaces=true,                 % keeps spaces in text, useful for keeping indentation of code (possibly needs columns=flexible)
  keywordstyle=\color{blue},       % keyword style
  language=Java,                 % the language of the code
  morekeywords={*,...},            % if you want to add more keywords to the set
  numbers=left,                    % where to put the line-numbers; possible values are (none, left, right)
  numbersep=5pt,                   % how far the line-numbers are from the code
  numberstyle=\scriptsize\color{black}, % the style that is used for the line-numbers
  rulecolor=\color{black},         % if not set, the frame-color may be changed on line-breaks within not-black text (e.g. comments (green here))
  showspaces=false,                % show spaces everywhere adding particular underscores; it overrides 'showstringspaces'
  showstringspaces=false,          % underline spaces within strings only
  showtabs=false,                  % show tabs within strings adding particular underscores
  stepnumber=1,                    % the step between two line-numbers. If it's 1, each line will be numbered
  stringstyle=\color{mymauve},     % string literal style
  tabsize=2,                       % sets default tabsize to 2 spaces
  title=\lstname,                  % show the filename of files included with \lstinputlisting; also try caption instead of title
  moredelim=[is][\bf]{*}{*},
}

%cachemiss full paper: atc18 
\section{Introduction}
In today's trusted execution environments (TEE), notably Intel software guard extensions or SGX, memory content is encrypted with the intent to protect data confidentiality against a privileged adversary (e.g., a malicious operating system).
However, recent practical attacks on SGX have demonstrated the feasibility of extracting the plain-text of memory content by observing only the memory access pattern disclosed through various micro-architectural side channels. For instance, the SGX enclave has to switch the control flow to the untrusted world to handle page faults, and a controlled side-channel attacker sitting in the untrusted world observes the sequence of page faults emitted from the enclave from which secret can be inferred. For another instance, various cache timing attacks can be mounted from the untrusted host machine. In these side-channels, the attack success requires the enclave to trigger page faults or cache misses. 

A promising design paradigm for mitigation is to leverage Intel's hardware transactional memory (TSX). TSX is an Intel CPU feature and co-exists with SGX. When it executes a program inside an TSX transaction, and cache misses or interruptions occur, the control is returned directly to the software without notifying operating systems. Therefore, the successful execution of a program inside an TSX transaction implies that the execution completes without any cache misses or interruptions (including page faults). In other words, the memory access pattern attacks that have to rely on cache misses or page faults would fail on the execution instance in an TSX transaction.

Existing researches~\cite{shih2017t,DBLP:conf/uss/GrussLSOHC17,DBLP:conf/ccs/ChenZRZ17} follow the above design paradigm and build systems/tools. Specifically, Cloak~\cite{DBLP:conf/uss/GrussLSOHC17} and D\'ej\`a Vu~\cite{DBLP:conf/ccs/ChenZRZ17} protect cryptographic libraries under cache-timing attacks. A typical cryptographic procedure has a small, cache-resident working set, and thus it affords to be directly mapped to a dedicated TSX transaction. T-SGX~\cite{shih2017t} supports more generic computation with working sets much larger than a CPU data cache. It partitions the target computation at the compilation time (as an LLVM extension), in such a way that each program partition has a working set small enough to fit into the CPU cache and thus can be run inside a TSX transaction. As a static scheme, T-SGX has to consider the worst-case at runtime and make each program partition conservatively small to avoid (deterministic) transaction aborts. For instance, on a $8$-way cache, each program partition from T-SGX should have no more than $8$ memory instructions. Otherwise, chances are that a transaction would request more memory accesses than a cache line of $8$ slots. If that occurs, a conflict cache miss is bound to happen, and the transaction is bound to abort.
Thus, transactions resulted from T-SGX or, more generally, from a static partitioning scheme, are rather small. Small transactions lead to finer-grained information leakage (degraded security) and higher performance overhead. Table~\ref{tab:distinction} summarizes existing researches in performance, side-channel security and data scalability (i.e., how large a working set can the target computation can support).

%where \cmark\kern-1.1ex\raisebox{.7ex}{\rotatebox[origin=c]{125}{--}} means T-SGX defends against certain attacks (page-fault controlled side-channels~\cite{DBLP:conf/sp/XuCP15}) but not other attacks, such as cache-timing based side-channel.

\begin{table}[!htbp] %force in current page, disable float.
\caption{Existing hardening works based on TSX and the distinction of this work.}
\label{tab:distinction}\centering{\footnotesize
\begin{tabularx}{0.49\textwidth}{ |c|p{1.5cm}|l|X|X|l| }
  \hline
Systems & Partitioning method & Scalability & Security against cache timing & Security against page faults & Perf.
  \\ \hline
Cloak~\cite{DBLP:conf/uss/GrussLSOHC17}
& 
\multirow{2}{*}{Manual} 
&
\multirow{2}{*}{\xmark} 
& 
\multirow{2}{*}{\cmark} %(Cache-timing attacks only)
& \multirow{2}{*}{\cmark} %(Cache-timing attacks only)
	& \multirow{2}{*}{\xmark}
  \\ \cline{1-1}
D\'ej\`a Vu~\cite{DBLP:conf/ccs/ChenZRZ17} 
	& & & & &\\ \hline
T-SGX~\cite{shih2017t} & Static & \cmark  & 
%\cmark\kern-1.1ex\raisebox{.7ex}{\rotatebox[origin=c]{125}{--}} %(Page-fault attacks only) 
\cmark & \xmark & \xmark
  \\ \hline
This work & Dynamic & \cmark & \cmark & \cmark & \cmark
  \\ \hline
\end{tabularx}
}
\end{table}

This work aims at scaling TSX-based enclave hardening to an arbitrarily large working set with strong security against memory-access pattern side channels and at low overhead.
Our insight is that a transaction's size is critically important to the performance and security of the transaction against side channels. A transaction too small leads to leaking fine-grained access patterns and high overhead per-instruction due to the limited degree of cost amortization. A transaction that is too large leads to deterministic abort and execution failure. See \S~\ref{sec:observation} for a preliminary study supporting our observation.
We propose to execute target computations in TSX transactions of proper sizes, that is, fully utilize the CPU cache. Ideally, one can enlarge a transaction to include many instructions as long as including the next instruction does not abort the transaction (e.g., deterministically causing cache misses). To achieve the goal, our approach is to {\it dynamically} insert transaction boundaries instead of doing so statically or manually as in the existing literature. By dynamic program partitioning, various runtime information, including current cache content and utilization, can be taken into account to predict the next cache miss, {\it more accurately}, and to result in larger transactions. The distinction of our research in comparison with existing TSX-based approaches is presented in Table~\ref{tab:distinction}.

We develop a C++ library to materialize the above design and support dynamic program partitioning. For supporting general computation, the library's API allows developers to specify data types of different access patterns in the target program. At runtime, the library partitions the target program and executes individual partitions in dedicated TSX transactions.
More specifically, before entering a TSX transaction, the library prepares the data layout in the main memory to map it to CPU caches with a small footprint.
When running inside the transaction, the library monitors all memory references (through the encapsulated data types), builds a cache model inside the CPU, and analyzes the cache utilization before landing the prediction of which instruction will cause the next cache miss. The prediction result assists the decision-making on when to insert a TSX instruction boundary (e.g., \texttt{xend}) without transaction aborts.
We carefully engineer the system to ensure the cache model has a small footprint in the CPU and does not interfere with the application data.

Using the developed library, we build an application to enable strongly secure data analytics in SGX enclave. We choose external oblivious algorithms to express the target data computation. Here, compared with the classic ``word''-oblivious algorithms (as adopted in Opaque~\cite{DBLP:conf/nsdi/ZhengDBPGS17} and ObliviousML~\cite{DBLP:conf/uss/OhrimenkoSFMNVC16}) and ORAMs~\cite{DBLP:conf/ccs/StefanovDSFRYD13} (as adopted in ZeroTrace~\cite{DBLP:journals/iacr/SasyGF17} and Prochlo~\cite{DBLP:conf/sosp/BittauEMMRLRKTS17}), external oblivious algorithms are a family of data oblivious algorithms with much lower time complexity and is specialized for common data analytical computations. 
An external oblivious algorithm accesses data in two places: 1) An external data source, the access to which does not leak secret information, and 2) an internal ``stash'' that needs to be protected from leakage.
We implement an external oblivious algorithm on our library by placing the external data source outside hardware transactions and placing the stash-accessing code inside transactions.
By this means, we implement a variety of data analytics for sorting, shuffle, binary search, $K$-means, etc.

%Altogether, the system achieves the notion of cache-miss obliviousness, that is, the trace of cache misses during enclave execution is oblivious to the memory data being accessed. This ensures strong security against most memory-access side-channels, including the physical-level bus tapping.

We evaluate the performance of the built data-analytical system by measuring its execution time.
%Specifically, we compare the transaction size resulted from dynamic partitioning with that from static partitioning (T-SGX) and manual partitioning schemes (Cloak).
We compare the execution time with that of word-oblivious algorithms (used in existing works, Opaque~\cite{DBLP:conf/nsdi/ZhengDBPGS17}). The result shows that our work effectively increases the transaction size and reduces the execution time, both by orders of magnitude.

This work makes the following contributions:

\vspace{2pt}\noindent$\bullet$\textit{
New technique}:
We propose a new technique for partitioning the program dynamically. We build a library to execute the partitioned programs in TSX transactions.
The library works by monitoring cache utilization and asserting a transaction's end upon a full cache.

\vspace{2pt}\noindent$\bullet$\textit{
Secure systems}:
We develop a data-analytical system with strong side-channel security. The system runs external oblivious algorithms on top of the library inside the SGX enclave. It achieves a new security notion, cache-miss obliviousness against all memory-access side channels.

\vspace{2pt}\noindent$\bullet$\textit{
Evaluation}:
We conduct a performance evaluation that shows our system can increase transactions' size and reduce execution time by orders of magnitude compared with the state-of-the-art solutions.

\section{Preliminaries}

\noindent{\bf Intel Software Guard eXtension (SGX)}: Intel SGX is an x86-64 ISA extension supported by Intel CPUs since the Skylake release in 2016. SGX enables isolated program execution from an otherwise untrusted host machine. At the hardware level, the SGX's secure world includes a tamper-proof CPU which automatically encrypts memory data in the enclave region upon cache-line write-back. Programs running outside the SGX's secure world can only read the cipher-text of enclave memory content. The enclave runs unprivileged program and excludes any OS kernel code, by explicitly prohibiting system services (e.g., system calls) inside an enclave. To use the technology, a client initializes an enclave by uploading the in-enclave program and uses SGX's seal and attestation mechanism~\cite{Anati_innovativetechnology} to verify the correct setup of the execution environment (e.g., by a digest of enclave memory content). During the program execution, the enclave is entered and exited voluntarily (by SGX instructions, e.g., \texttt{EENTER} and \texttt{EEXIT}) or passively (by interrupts or traps). These world-switch events trigger the context saving/loading in both hardware and software levels. Comparing prior TEE solutions~\cite{me:txt,me:tpm,me:tzone,me:scpu}, SGX uniquely supports multi-core concurrent execution, dynamic paging, and interrupted execution.

\noindent{\bf Intel Transactional Synchronization eXtension (TSX)}: TSX is an Intel CPU ISA extension designed to enable atomic execution of a program. A program is supposed to be executed atomically in the unit of so-called transaction. Any dirty cache line is not written back until the commit of a transaction. A TSX transaction aborts under various causes: It aborts a transaction when data conflict is detected at the transaction commit time (Abort Cause 1). In order to detect data conflict, the hardware keeps track of a transaction's readset and writeset. The writeset needs to be inside the L1 data cache (L1D) and readset inside the L3 cache. Thus, it aborts the transaction when dirty data-cache lines are evicted, triggering cache write-back, before the end of transaction (Cause 2). It also aborts the transaction when the readset exceeds the L3 cache (Cause 3). In addition, it aborts a transaction upon various systems events such as page-fault, interrupts and other exceptions delivered to the processor (Cause 4).

\section{Threat Model}
\label{sec:threat}

\noindent
This work considers the practice of outsourced computations on untrusted hosts. For instance, a data owner outsources their security-sensitive computations to a third-party host, such as a public cloud service, and let the computation be executed there. 
The host machines are with Intel CPUs supporting both SGX and TSX. The data owner deploys their program in the SGX enclave and upload their data to the host machine encrypted but accessible by the enclave. The data owner trusts the CPU at the hardware layer of the host and the owner-attested program at the software layer. Hardware and software outside this perimeter of the enclave is untrusted by the data owner. 

\noindent{\bf Threats}: 
This work considers the memory-access pattern attacks in which the attacker controls privileged software outside the enclave including operating systems. Specifically, the attacker monitors the trace of micro-architectural events emitted from the victim enclave execution and, based on them, further infer sensitive information. There are two particular side channels, that is, controlled side-channel attacks monitoring the page faults~\cite{DBLP:conf/sp/XuCP15} and cache-timing attacks monitoring cache misses (e.g., the cache misses on page-table access/dirty bits~\cite{DBLP:conf/uss/0001SGKKP17,DBLP:conf/uss/BulckWKPS17} an others~\cite{DBLP:conf/ccs/WangCPZWBTG17,DBLP:conf/usenix/HahnelCP17}). 

%Access-driven attacks against data-intensive computations can lead to significant information disclosure~\cite{DBLP:conf/ccs/OhrimenkoCFGKS15,DBLP:conf/nsdi/ZhengDBPGS17,DBLP:journals/corr/BaterEEGKD16}.

Note that this work does not mitigate other side-channel attacks including execution timing attacks, power analysis, denial-of-service attacks, rollback attacks, etc. Existing defensive techniques can be applied on top of our system to mitigate these attack vectors.

\noindent{\bf Security goals}:
Given an execution instance $I$ inside the enclave, the working set consists of two types of data, the data whose access trace leaks the value and the data whose access trace does not leak. This work focuses on achieving the following tractable security goals: 1) when $I$'s access-leaky data is smaller than a CPU data cache, our system named by CMO should ensure that the trace of any micro-architectural events emitted from $I$ is oblivious to the value in $I$'s working set. Informally, the attacker capable of monitoring the micro-architectural side channels cannot distinguish the case that $I$'s working set contains a specific value from the case that it does not. This is the notion called cache-miss obliviousness. 2) when $I$'s access-leaky data is much larger than a CPU data cache, our system ensures that the information leaked about the value of access-leaky data from the side channels occurs at the granularity of transactions. 

\section{Design Motivation}
\label{sec:observation}

\begin{figure}
\centering
\includegraphics[width=0.35\textwidth]{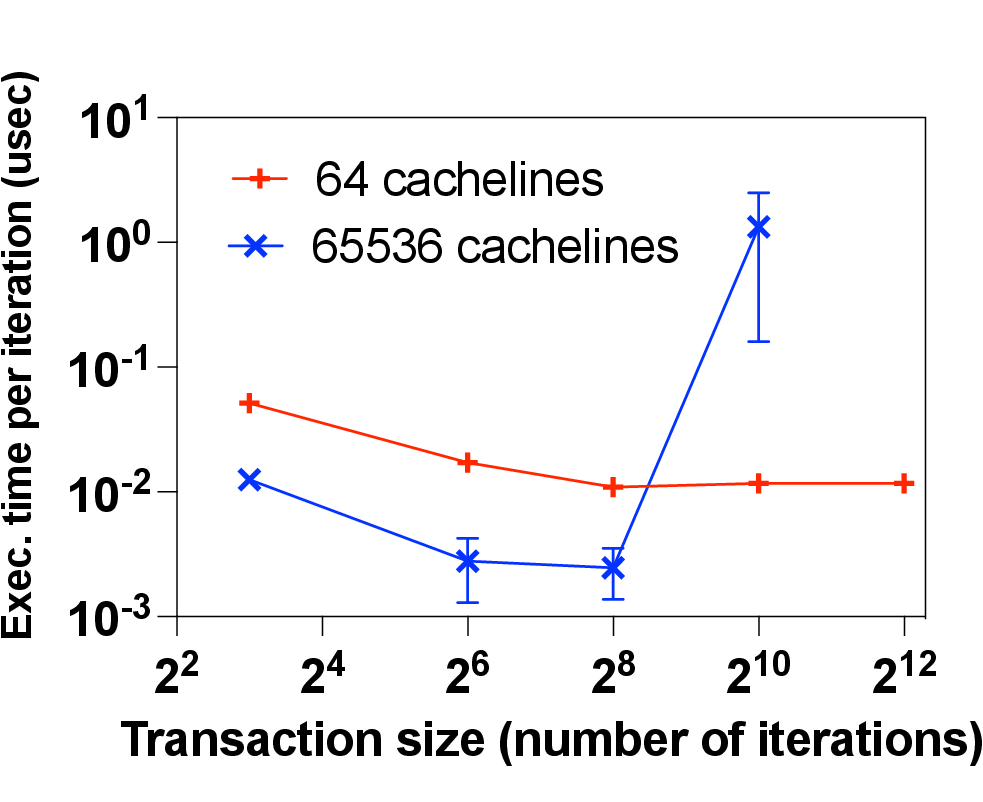}
\caption{Preliminary performance observation with varying transaction sizes (with two arrays respectively of $64$ and $65536$ elements/cachelines).}
\label{exp:txsize:prel}
\end{figure}

This preliminary measurement study is designed to observe how the per-instruction execution time changes with the size of TSX transactions. We write a simple C++ program that writes to an array in a loop. Each array element has the same size with the cache line of the machine where the experiment is conducted. By this means, each access to an element can lead to a cache miss. The size of a transaction measures the number of instructions/iterations in the loop. Given a transaction size $t$, we wrap $t$ iterations in a TSX transaction by inserting \texttt{xend} and \texttt{xbegin}. We run the program with an array of 64 elements on an Intel SGX machine with the following specs: Intel(R) Core(TM) i5-7500 CPU @ 3.40GHz 64 bits, with 32 KB L1 cache size (8-way 64 sets) and 8 GB RAM. 
We consider two settings of array length: $64$ elements and $65536$ elements.
In the experiment, we vary the transaction size and report the average execution time per instruction over 1000 runs of the experiment. We also report the standard deviation.

The result in Figure~\ref{exp:txsize:prel} shows that the execution time generally decreases as transactions grow larger. Consider, first, the result with the $64$-element array.
When the transaction size increases from $8$ to $64$ iterations, the execution time decreases by $2\sim{}5\times$. This can be explained by that with larger transactions, the TSX transaction execution overhead (e.g., running \texttt{xbegin}/\texttt{xend} instructions and possible aborts) can be amortized to more instructions, leading to shorter execution time per instruction.
Note that transaction size $8$ is the maximal transaction size that can be set by a static scheme, such as T-SGX, on a CPU of $8$-way cache.

With the array of $64$ elements/cachelines, a transaction of maximal size would include all instructions updating the $64$ cachelines, does not cause cache misses and thus does not abort.
By contrast, this is not the case with the larger array of $65536$ elements/cachelines: A transaction of maximal size would cause cache misses and deterministically abort. As shown in Figure~\ref{exp:txsize:prel}, with the array of $65536$ elements, when the transaction is too large (e.g., 1024 loop iterations), it would incur very high overhead. Further when the transaction size is increased to $4096$ iterations, the transaction winds up in endless abort-re-execution and cannot complete. 

This preliminary study implies that while the transaction size resulted from static partitioning is too small and incurs high amortized overhead per instruction, enlarging transaction too much is not a good idea either. A proper transaction size that leads to good performance is the medium one, such as a transaction of $256$ iterations for an array of $65536$ elements in Figure~\ref{exp:txsize:prel}. In this work, we quantify this intuition by enlarging a TSX transaction until the next instruction to include would deterministically cause cache misses.

\section{Library for Dynamic Program Partitioning}

\begin{figure} 
\centering
\includegraphics[width=0.225\textwidth]{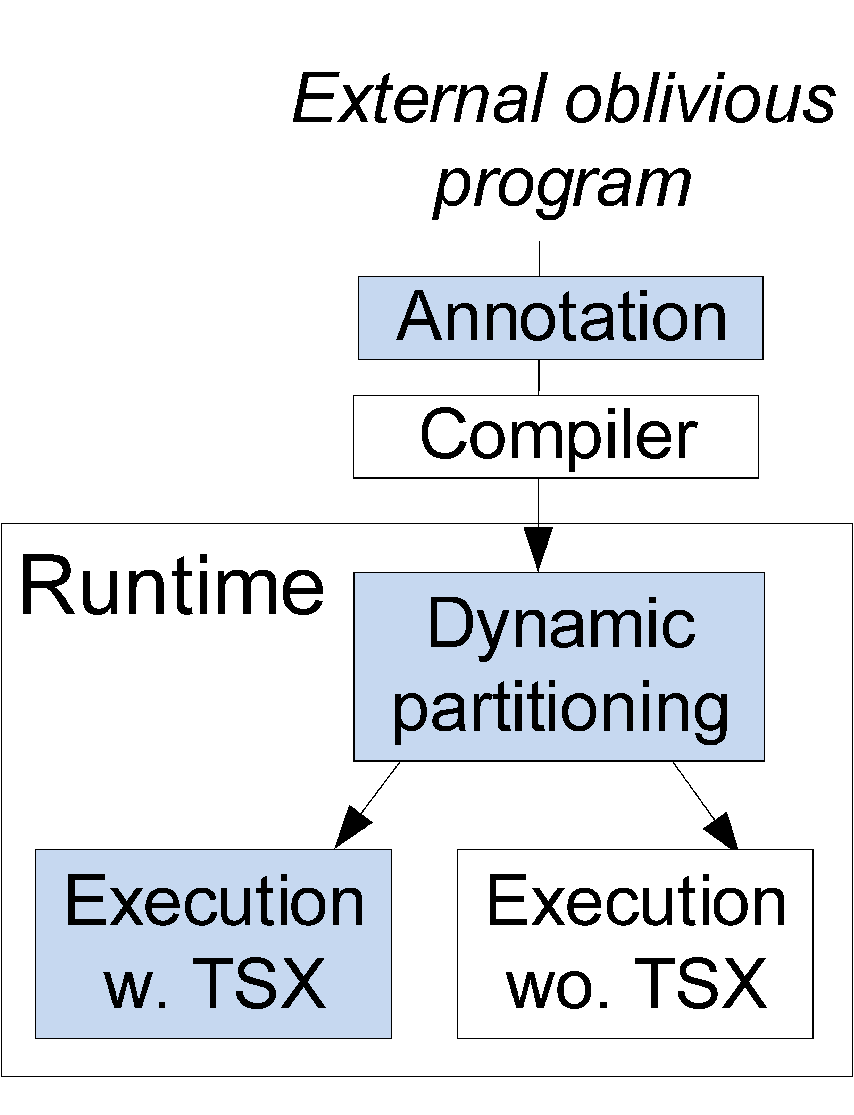}
\caption{System overview: A target computation is annotated by our programming API that specifies data-access pattern. Our runtime system first partitions the program dynamically to units (i.e., dynamic partitioning), and then runs each unit in a TSX transaction (i.e., tx-execution engine).}
\label{fig:overview}
\end{figure}

We design and develop a runtime system enforcing external data obliviousness inside SGX enclaves leveraging Intel TSX features.
Given a program expressing an external oblivious algorithm, the system provides an API (S1) for developers to annotate the leaky section of the program. Given the annotated program, we provide an execution engine (S2) that runs the program on the architecture of two execution environments: the one with access-pattern protection (e.g., prohibiting cache-misses) and the other without. We call the former by fortified enclave and the latter by unprotected enclave. Briefly, the execution engine partitions the runtime instance dynamically in a way that assures no cache-miss yet can scale to relatively large data.
Figure~\ref{fig:overview} illustrates the overview of our system. In the following, we present the detailed design and implementation of our system in each component, S1 in \S~\ref{sec:api} and S2 in \S~\ref{sec:dynpart}.

\subsection{Annotating Target Programs}
\noindent
\label{sec:api}

%{\bf API}: 
Our system provides a programming interface for C++ developers to annotate the leaky section of the program and define the type of data referenced inside the leaky section. Specifically, 1) we provide two library functions, \texttt{begin\_leaky()} and \texttt{end\_leaky()}, for developers to declare the begin and end of a program snippet where leaky memory access occurs. 2) In addition, all the memory data accessed inside a leaky section needs to be defined with our data type. We provide four data-container types as defined below. The explicit data type allows our execution engine to know ahead of time the possible data-access pattern and take action in partitioning more precisely.

The data-container types are classified based on whether the access is leaky and whether the data is read-only. The four data types are \texttt{NobRO}, \texttt{ObRO}, \texttt{NobRW}, \texttt{ObRW}, where \texttt{ObRW} stands for oblivious read-write data container and NobRO is non-oblivious read-only data container. The API of the data classes are provided in Figure~\ref{lst:samplecode}. We also provide an example code that declares the merge passes in a merge sort algorithm to be leaky sections.

\definecolor{mygreen}{rgb}{0,0.6,0}
\begin{figure}[h!]
%\begin{scriptsize}
%\begin{tabular}{cc}
\begin{minipage}{.5\textwidth}
\lstset{ %
  xleftmargin=3.0ex,
  backgroundcolor=\color{white},   % choose the background color; you must add \usepackage{color} or \usepackage{xcolor}
  %basicstyle=\footnotesize\ttfamily,        % the size of the fonts that are used for the code
  basicstyle=\small\ttfamily,        % the size of the fonts that are used for the code
  breakatwhitespace=false,         % sets if automatic breaks should only happen at whitespace
  breaklines=true,                 % sets automatic line breaking
  captionpos=b,                    % sets the caption-position to bottom
  commentstyle=\color{mygreen},    % comment style
  deletekeywords={...},            % if you want to delete keywords from the given language
  escapeinside={\%*}{*)},          % if you want to add LaTeX within your code
  extendedchars=true,              % lets you use non-ASCII characters; for 8-bits encodings only, does not work with UTF-8
  %frame=single,                    % adds a frame around the code
  keepspaces=true,                 % keeps spaces in text, useful for keeping indentation of code (possibly needs columns=flexible)
  keywordstyle=\color{blue},       % keyword style
  language=Java,                 % the language of the code
  morekeywords={*,...},            % if you want to add more keywords to the set
  numbers=left,                    % where to put the line-numbers; possible values are (none, left, right)
  numbersep=5pt,                   % how far the line-numbers are from the code
  numberstyle=\small\color{black}, % the style that is used for the line-numbers
  rulecolor=\color{black},         % if not set, the frame-color may be changed on line-breaks within not-black text (e.g.n comments (green here))
  showspaces=false,                % show spaces everywhere adding particular underscores; it overrides 'showstringspaces'
  showstringspaces=false,          % underline spaces within strings only
  showtabs=false,                  % show tabs within strings adding particular underscores
  stepnumber=1,                    % the step between two line-numbers. If it's 1, each line will be numbered
  stringstyle=\color{mymauve},     % string literal style
  tabsize=2,                       % sets default tabsize to 2 spaces
  title=\lstname,                  % show the filename of files included with \lstinputlisting; also try caption instead of title
  moredelim=[is][\bf]{*}{*},
}
\begin{lstlisting}
class ObRO {
  int32_t read_next(); 
  void reset();
}
class ObRW {
  void write_next(int32_t data); 
  void reset();
}
class NobRW {
  int32_t read_at(int32_t addr);
  void write_at(int32_t addr, int32_t data); 
}
class NobRO {
  int32_t nob_read_at(int32_t addr); 
}
\end{lstlisting}
\end{minipage}
\\
\vspace{-0.35in}
\begin{minipage}{.5\textwidth}
\lstset{ %
  xleftmargin=3.0ex,
  backgroundcolor=\color{white},   % choose the background color; you must add \usepackage{color} or \usepackage{xcolor}
  %basicstyle=\footnotesize\ttfamily,        % the size of the fonts that are used for the code
  basicstyle=\small\ttfamily,        % the size of the fonts that are used for the code
  breakatwhitespace=false,         % sets if automatic breaks should only happen at whitespace
  breaklines=true,                 % sets automatic line breaking
  captionpos=b,                    % sets the caption-position to bottom
  commentstyle=\color{mygreen},    % comment style
  deletekeywords={...},            % if you want to delete keywords from the given language
  escapeinside={\%*}{*)},          % if you want to add LaTeX within your code
  extendedchars=true,              % lets you use non-ASCII characters; for 8-bits encodings only, does not work with UTF-8
  %frame=single,                    % adds a frame around the code
  keepspaces=true,                 % keeps spaces in text, useful for keeping indentation of code (possibly needs columns=flexible)
  keywordstyle=\color{blue},       % keyword style
  language=Java,                 % the language of the code
  morekeywords={*,...},            % if you want to add more keywords to the set
  numbers=left,                    % where to put the line-numbers; possible values are (none, left, right)
  numbersep=5pt,                   % how far the line-numbers are from the code
  numberstyle=\small\color{black}, % the style that is used for the line-numbers
  rulecolor=\color{black},         % if not set, the frame-color may be changed on line-breaks within not-black text (e.g. comments (green here))
  showspaces=false,                % show spaces everywhere adding particular underscores; it overrides 'showstringspaces'
  showstringspaces=false,          % underline spaces within strings only
  showtabs=false,                  % show tabs within strings adding particular underscores
  stepnumber=1,                    % the step between two line-numbers. If it's 1, each line will be numbered
  stringstyle=\color{mymauve},     % string literal style
  tabsize=2,                       % sets default tabsize to 2 spaces
  title=\lstname,                  % show the filename of files included with \lstinputlisting; also try caption instead of title
  moredelim=[is][\bf]{*}{*},
}
\begin{lstlisting}
void MergeSort(int[] array, int start, int end){ 
  if (start == end - 1) return;
  MergeSort(array, start, (end + start)/2); 
  MergeSort(array, (end + start)/2, end);
  Merge(array, 0, (end + start)/2, (end + start)/2, end);
}
void Merge(int[] array, int start1, int end1, int start2, int end2){
  *NobRW* parray = new *NobRW*(array);
  *begin_leaky();*
  originalMerge(parray, start1, end1, start2, end2); 
  *end_leaky();*
}
\end{lstlisting}
\end{minipage}
%\end{tabular}
%\end{scriptsize}
\caption{Implementing merge sort based on the CMO API: in-transaction data types and leaky section}
\label{lst:samplecode}
\end{figure}

\subsection{Realizing Dynamic Program Partitioning}
\label{sec:dynpart}

\noindent{\bf Design motivation}:
Existing side-channel defenses leverage the hardware transaction features to detect any unintended cache misses. Notably, Cloak~\cite{DBLP:conf/uss/GrussLSOHC17} detects cache attacks using Intel TSX. Briefly, it executes the computation in HTM transactions, such that the intra-transaction memory accesses are resolved by cache hits and thus are concealed from the adversary. Note that the inter-transaction memory accesses are disclosed. The security of these Cloak alike schemes requires that access-leaky memory regions must reside in data cache; this security requirement significantly limits the data scale. In addition, Cloak's programming interface is rather low-level and requires the programmer to manually specify the scope of individual HTM transactions, which further limits its applicability in big-data computations.

To systematically improve the data scalability, we study two technical problems: (O1) dynamic program partitioning and (O2) enlarging transaction. First, given a target computation, we dynamically partition the program to smaller execution units, each run in an individual transaction. Compared with existing work, such as Cloak~\cite{DBLP:conf/uss/GrussLSOHC17} and T-SGX~\cite{shih2017t}, our unique perspective about program partitioning is to take into account various runtime information.

Second, for each execution unit, we enlarge the transaction as much as possible. Here, the larger a transaction is, the more performance efficient it can be, as the transaction setup cost (e.g., running \texttt{xbegin}/\texttt{xend} instructions) can be amortized among more instructions. Note that the size of any TSX transaction has a theoretic limit in that its working set cannot exceed a CPU data cache. Our goal is to reach this theoretic bound in transaction size. 

\noindent{\bf 
Design rationale}: 
As revealed in our API design in \S~\ref{sec:api}, a starting point of our approach is to expose to the runtime system some high-level semantic information. This information includes different types of data accesses; concretely, the data-container types in our API lets the runtime be aware of four data-access classes: 1) non-oblivious reads (or access-leaky reads as in \texttt{NobRO}), 2) non-oblivious writes (or access-leaky writes as in \texttt{NobRW}), 3) oblivious reads (as in \texttt{ObRO}) and 4) oblivious writes (as in \texttt{ObRW}). 

\begin{figure} 
\centering
\includegraphics[width=0.35\textwidth]{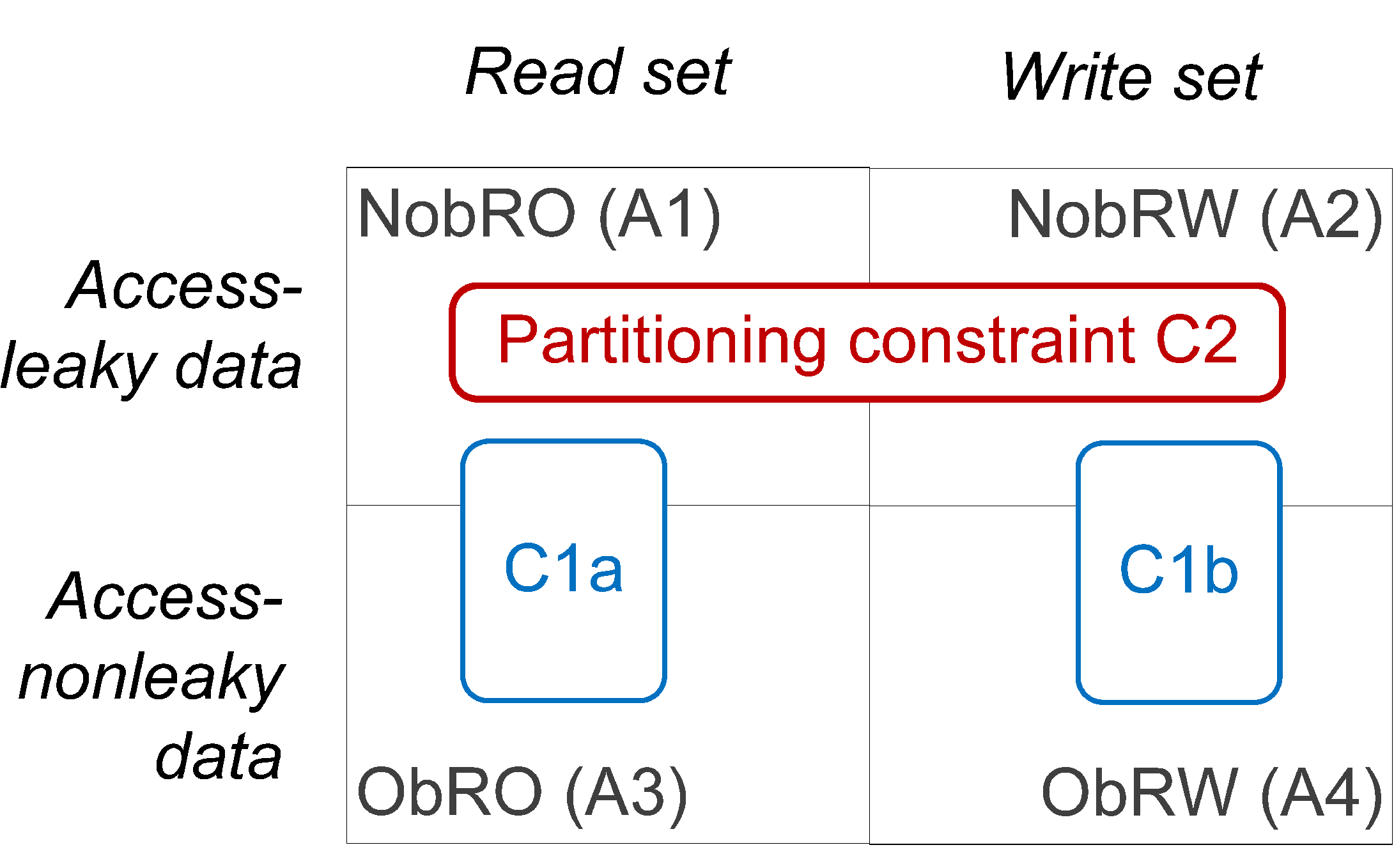}
	\caption{Constraints in partitioning programs to TSX transactions: Constraints are imposed by the requirement of hardware (C1a and C1b) and security protection (C2). Data is classified by read-only or read-write in transactions and by whether its access leaks sensitive information.} 
\label{fig:constraints} 
\end{figure}

With the awareness of data access type, we can specify the constraints of program partitioning problem. In general, there are two constraints, the constraint imposed by the TSX hardware (C1) and that imposed by the security requirement (C2). 
For C1, it requires that a transaction's write-set should not cause L1 cache replacement, and a transaction's read-set should not cause replacement in the last-level cache (LLC).
For C2, the access-leaky memory region must reside in data cache. Here, an access-leaky memory region is defined by memory data accessed in a data-dependent way; for instance, \texttt{NobRO} and \texttt{NobRW} instances are access-leaky regions.\footnote{We may use terms \texttt{NobXX} to represent access-leaky regions}. The three constraints are summarized in the list below. Also, Figure~\ref{fig:constraints} illustrates the three constraints in the presence of four types of data accesses. 

\begin{itemize}
\item C1a): Make transaction readset smaller than CPU LLC (last level cache).
\item C1b): Make transaction writeset smaller than CPU L1 cache.
\item C2): Keep access-leaky data inside the CPU data cache.
\end{itemize}

%The constraint of C2 is due to the strong security concern. Because the HTM based protection only conceals the intra-transaction memory access pattern, and the inter-transaction memory access pattern is disclosed. Placing the leaky memory accesses across multiple transactions would inevitably disclose the data secrets. 

%Because of the constraints (particularly C2), we focus on {\it partitioning the non-leaky part of the program}. That is, we consider the problem of partitioning the program into smaller units, each unit accessing a partition of non-leaky data (\texttt{ObXX}) but accessing leaky data in their entirety. Note that the leaky data is not partitioned and has to be loaded in total to the data cache. This design choice implies that we only support computation with small cache-bound leaky data (\texttt{NobXX}).\footnote{Note that the assumption presents a fundamental limitation of our approach in supporting computation of large access-leaky data.}, which is consistent with our target computations (as described in \S~\ref{sec:computation}).  

Our goal is two-fold: (O1) Support computations of (non-leaky) data as large as possible by dynamically partitioning the computations, 
and (O2) Run these computations using as large transactions as possible.

\begin{figure} 
  \centering
  \includegraphics[width=0.475\textwidth]{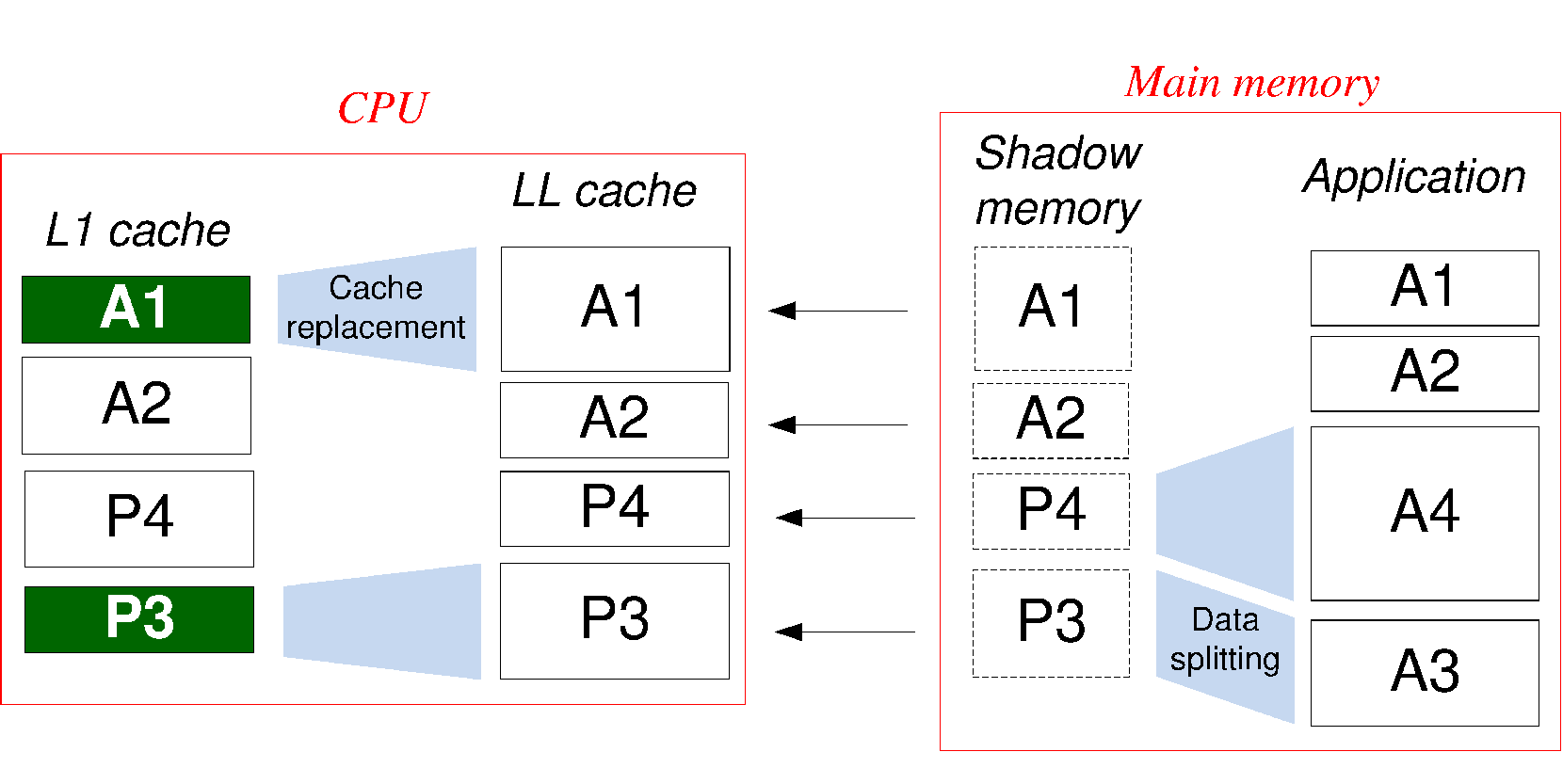} 
  \caption{Splitting data to transaction-wise partitions and mapping data to the memory/cache hierarchy: Shadow-memory is an indirection layer that arranges data in cache based on the access pattern. From LL (Last-Level) cache to L1 (Level-one) cache, read-only data replacement is enabled. In the figure, A1 (A2) represents the read-only (read-write) data whose access pattern leaks sensitive information, and A3/P3 (A4/P4) represents the read-only (read-write) data whose access pattern does not leak sensitive information.} 
\label{fig:shadow} 
\end{figure}

The high-level approach is illustrated in Figure~\ref{fig:shadow}. As in the figure, consider a computation whose application memory consists of (cache-bound) access-leaky data (i.e., A1 and A2) and large non-leaky data (i.e., A3 and A4). We split the non-leaky data into a series of smaller data partitions (i.e., \{P3\} and \{P4\}), each of which is small enough to fit in a CPU data cache. 
To support large transactions (O2), we enable {\it L1 cache replacement on read-only data}. That is, the footprint of A1 and P3 in L1 cache can be reduced to just one set, as will be explained next.

\fbox{\parbox{0.85\linewidth}{
Formally, our work can be formulated as an optimization problem: Given a program $G$ on data $\{A1,A2,A3,A4\}$, it partitions $G$ to a series of $\{g\}$ with each program partition $g$ operates on data $A1,A2,P3,P4$ such that it maximizes $\min{(P3,P4)}$ and it is subject to the following constraints:

\begin{eqnarray}
  \nonumber
\cup\{P3\} & = & A3\\
  \nonumber
\cup\{P4\} & = & A4\\
  \nonumber
A1 + P3 + A2 + P4 & < & LLC\\
  \nonumber
A1/L + P3/L + A2 + P4 & < & L1
\end{eqnarray}

Here, $L$ is the maximal number of LLC cache lines that can be mapped to one L1 cache set.
}}

\label{sec:shadowmem}
\noindent{\bf Large transactions by shadow memory}:
As will be described next, read-only data (A1 and P3) in a transaction can reuse L1 data cache without aborting the transaction. 

%cache full utilization is essential.\footnote{For instance, a poorly utilized cache may abort a transaction when just one set (out of say 64 cache sets) is full. A well-utilized cache may abort a transaction when the entire cache is full, which results in a much larger transaction.} 

Efficient use of the cache is of particular importance to ensuring large transactions. The transaction will not abort when the cache is underutilized; for instance, the transaction may end when only one cache set (out of 64) becomes full while the other sets are empty.

We propose a technique called shadow memory that arranges the in- transaction memory layout such that memory data is mapped to the CPU cache hierarchy without unnecessary conflict. Briefly, we enforce the rules that different types of data (A1, ...A4) should not overlap in data caches. Specifically, consider a cache hierarchy with a 8-way 32KB L1 cache and 16-way 8MB L2 cache. We allocate $60$ L1C sets to be A2 (\texttt{NobRW}), $2$ L1C sets to be P4, $1$ L1C set to be P3 and $1$ L1C set to be A1. Note that A1 and P3 can reuse the L1 cache and their sizes are bounded by LLC. Figure~\ref{fig:shadow} illustrates an example of shadow memory.

\noindent
{\bf Implementation notes}:
When realizing the dynamic partitioning, we encountered two design problems: 1) How and when to insert the \texttt{xbegin} and \texttt{xend} instructions to properly declare transaction boundaries such that the transactions can successfully finish their execution (i.e., without abort) yet are not prematurely ended? 2) How to realize the shadow memory in its full life cycle? 3) How to decide the condition that the computation cannot be executed using transactions at an early time (e.g., when the A1 is larger than LLC).

\begin{figure} 
  \centering
  \includegraphics[width=0.4\textwidth]{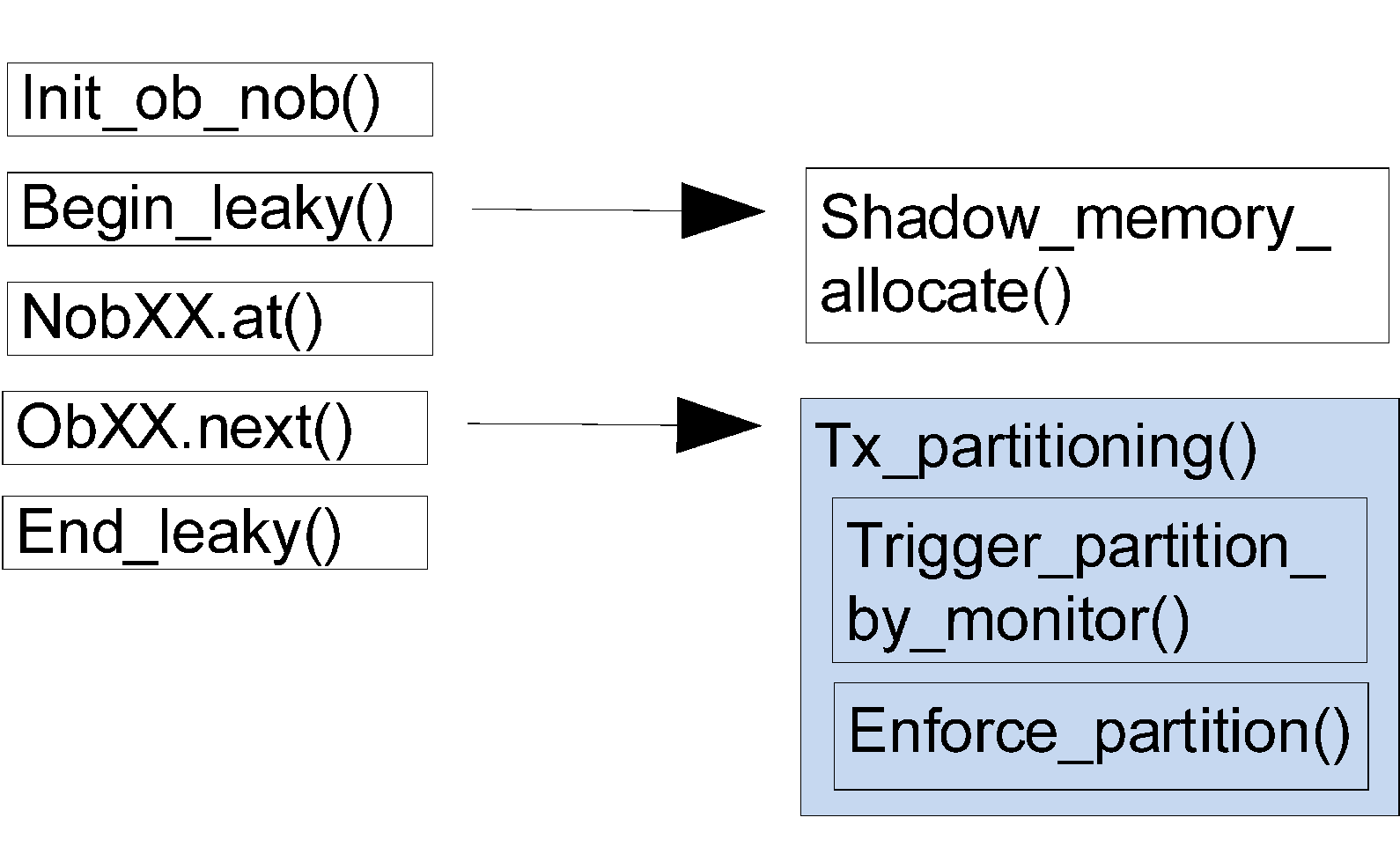}
\caption{Library implementation: Each box represents a function in our library. White boxes run outside TSX transactions and blue boxes run inside TSX transactions.} 
\label{fig:libimpl} 
\end{figure}

We build a library that realizes the partitioning life cycle. Recall that our API supports declaring the leak section's scope and wrapping each in-transaction memory access by custom data types. Upon leaky section declaration, we realize the allocation of shadow memory. Upon the in-transaction memory access, we hook the transaction partitioning schemes. The overall mechanism is illustrated in Figure~\ref{fig:libimpl}, and pseudo-code is presented in Listing~\ref{lst:samplecode:hook}.

Concretely, for shadow memory allocation, we allocate the shadow memory according to the layout described in \S~\ref{sec:shadowmem}. Note that the shadow-memory size is fixed (as the total size of a cache is fixed). We perform checks on whether the data size, specifically the size of access-leaky memory data, is small enough to fit into the data cache. In other words, we enforce the constraint C1 that A1 does not exceed L1 and A2 does not exceed LLC.

For transaction partitioning, we implement two components: 1) monitoring cache usage, and 2) realizing transaction partitioning. For 1), we simply monitor the condition that if each non-leaky data overflows, that is, the iterator call to \texttt{ObXX.next()}\footnote{\texttt{ObXX} refers to non-leaky data type, either \texttt{ObRO} or \texttt{ObRW}.} reaches its full capability. Note that we enforce a joint capability limit, that is, the total size of \texttt{ObXX} objects is bounded instead of individual \texttt{ObXX}.

For 2), we insert the TSX instructions \texttt{xend} and \texttt{xbegin}. Between them, we make data copy between shadow memory and application memory of the non-leaky data. Because the computation is partitioned based on non-leaky data, different transactions in the same leaky-section computation will use different non-leaky data, thus making it necessary to reload non-leaky data across transactions. Also, we preload data twice across transactional boundary.

\definecolor{mygreen}{rgb}{0,0.6,0}
\lstset{ %
  xleftmargin=3.0ex,
  backgroundcolor=\color{white},   % choose the background color; you must add \usepackage{color} or \usepackage{xcolor}
  %basicstyle=\footnotesize\ttfamily,        % the size of the fonts that are used for the code
  basicstyle=\small\ttfamily,        % the size of the fonts that are used for the code
  breakatwhitespace=false,         % sets if automatic breaks should only happen at whitespace
  breaklines=true,                 % sets automatic line breaking
  captionpos=b,                    % sets the caption-position to bottom
  commentstyle=\color{mygreen},    % comment style
  deletekeywords={...},            % if you want to delete keywords from the given language
  escapeinside={\%*}{*)},          % if you want to add LaTeX within your code
  extendedchars=true,              % lets you use non-ASCII characters; for 8-bits encodings only, does not work with UTF-8
  %frame=single,                    % adds a frame around the code
  keepspaces=true,                 % keeps spaces in text, useful for keeping indentation of code (possibly needs columns=flexible)
  keywordstyle=\color{blue},       % keyword style
  language=Java,                 % the language of the code
  morekeywords={*,...},            % if you want to add more keywords to the set
  numbers=left,                    % where to put the line-numbers; possible values are (none, left, right)
  numbersep=5pt,                   % how far the line-numbers are from the code
  numberstyle=\small\color{black}, % the style that is used for the line-numbers
  rulecolor=\color{black},         % if not set, the frame-color may be changed on line-breaks within not-black text (e.g. comments (green here))
  showspaces=false,                % show spaces everywhere adding particular underscores; it overrides 'showstringspaces'
  showstringspaces=false,          % underline spaces within strings only
  showtabs=false,                  % show tabs within strings adding particular underscores
  stepnumber=1,                    % the step between two line-numbers. If it's 1, each line will be numbered
  stringstyle=\color{mymauve},     % string literal style
  tabsize=2,                       % sets default tabsize to 2 spaces
  title=\lstname,                  % show the filename of files included with \lstinputlisting; also try caption instead of title
  caption={Library implementation: Hook partitioning triggers in non-leaky data scans.},                  % show the filename of files included with \lstinputlisting; also try caption instead of title
  label={lst:samplecode:hook},
  moredelim=[is][\bf]{*}{*},
}
\begin{lstlisting}
begin_leaky(){
  shadow_memory_alloc();
}
//Dynamically partition computations in next();
ObXX::next(){
  //1. monitor the overflow
  if(shadow_mem.ob.size + 1 < shadow_mem.ob.capability()) return;
  //2. start to partition
  if(!first_time) xend();
  //reload non-leaky data (ob) in shadow memory from application memory
  shadow_mem.reload_ob();
  double_preload();
  xbegin();
}
\end{lstlisting}

\subsection{Security Analysis}

Suppose running a target program in CMO under the threats of controlled side-channel and cache-timing attacks. We analyze the trace of page faults and cache misses during the program execution.

Note that the successful completion of an TSX transaction ensures no cache misses or page faults occur inside the transaction. Thus, in the trace, cache misses or page faults could only occur 1) between a successful transaction and the next transaction, or 2) at the end of an aborted transaction. For Case 1), the cache misses or page faults are caused by reloading the shadow memory in CMO. Because the shadow memory is reloaded by sequential data accesses, the cache misses or page faults do not leak information the data-element granularity.   More specifically, if it is to reload the shadow memory with an obliviously-accessed object (ObXX), the leaked access information does not disclose any correlation to the data value. If it is to reload the shadow memory with a non-obliviously accessed object (NobXX), the leaked access information disclose the data value at transaction granularity. 

%For instance,  XXX

For Case 2), transaction aborts that do occur in CMO cannot be caused by the cache misses predicated by our scheme. Thus, transactions cannot be aborted by conflict or capacity cache misses, as our cache model can precisely predict both causes. In addition, there is no compulsive cache miss, assuming the comprehensive shadow-memory reloading. CMO transactions can only be aborted due to interruptions raised by hardware or operating systems in a non-deterministic fashion.

We clarify this work does not handle execution-timing side channels; existing constant-time algorithms and designs can be applied on top of our system to handle the execution timing side-channels. 

\section{Building Side-Channel Secure Data Analytics}
\label{sec:cases}
\label{sec:computation}

While our library supports generic computation, we use it to implement the external-oblivious algorithms as a prototype system of secure data analytics.
We first present the preliminary of external oblivious algorithms and then describe how to implement them using our library.

\subsection{Preliminary: External Data-Oblivious Algorithms}
\label{sec:prel:externalobl}

External data-oblivious algorithms are a class of oblivious mechanisms that have advantageous time complexity at the expense of assuming a trusted internal memory or so-called ``stash''. External oblivious algorithms are proposed for a wide variety of computations, including shuffle (Melbourne shuffle~\cite{DBLP:conf/icalp/OhrimenkoGTU14}), sort (oblivious merge sort~\cite{DBLP:conf/ndss/WilliamsS08}), most aggregation computations, etc. Compared with word-oblivious algorithms (e.g., sorting networks causing a multiplicative factor of $O(\log{N})$), an external oblivious algorithm has better time complexity (e.g., Melbourne shuffle based sorting~\cite{DBLP:conf/ccs/OhrimenkoCFGKS15,DBLP:journals/popets/DangDCO17} with the O(1) multiplicative factor). Compared with ORAM (with a multiplicative factor of $O(\log^2{N})$), an external oblivious algorithm is specific to target computation and is more efficient.
While external oblivious algorithms achieve better complexity, they may present limitations on data scalability. Concretely, most existing external oblivious algorithms are designed for a client-server setting and assume a large internal memory at the client side. For instance, a Melbourne shuffle assumes an internal memory of $O(\sqrt{N})$. The large internal memory may be suited for a client machine (in the big-data setting), but presents challenges when treating cache as internal memory, whose size is very limited; for instance, a L1 cache is 32 KB and L3 cache is 8 MB in an Intel SGX CPU. Very recently, space-efficient external oblivious algorithms are proposed, for instance, stash shuffle with small $O(\log{N})$ internal space.

\subsection{Implementing External Oblivious Data Analytics}

Here, we present our experience building various external-oblivious computations using our library. 

\noindent{\bf $K$-means}:
A $K$-means computation takes as input a data array and randomly initialized K centroids. It produces the output of K centroids in K clusters that are the most representative of the data array. Here, each element in the data array represents a data point, and there are pair-wise distances defined between them. The $K$-means computation runs iteratively, where each iteration consists of two data passes, one (KM1) is a nested loop that (re)-assigns every array element to the closest centroid, and the other (KM2) is an array scan that randomly accesses the centroid array to update the centroid position. 

K-means is an external oblivious computation in the sense that it keeps the centroid as internal memory accessed in a leaky way and the data array as an external memory, which is accessed obliviously. 

To implement $K$-means using our library, we declare the data pass (KM2) to be the leaky section where the internal memory is accessed. In the leaky section, the centroid array is defined as a NobRW, data array as an \texttt{ObRO}, the data-centroid mapping as an \texttt{ObRW}. 
%The related sample code is in Listing~\ref{XXX}.

\begin{figure*}[!tbhp]
\centering
\subfloat[Binary search: Number of queries]{%
  \includegraphics[width=0.25\textwidth]{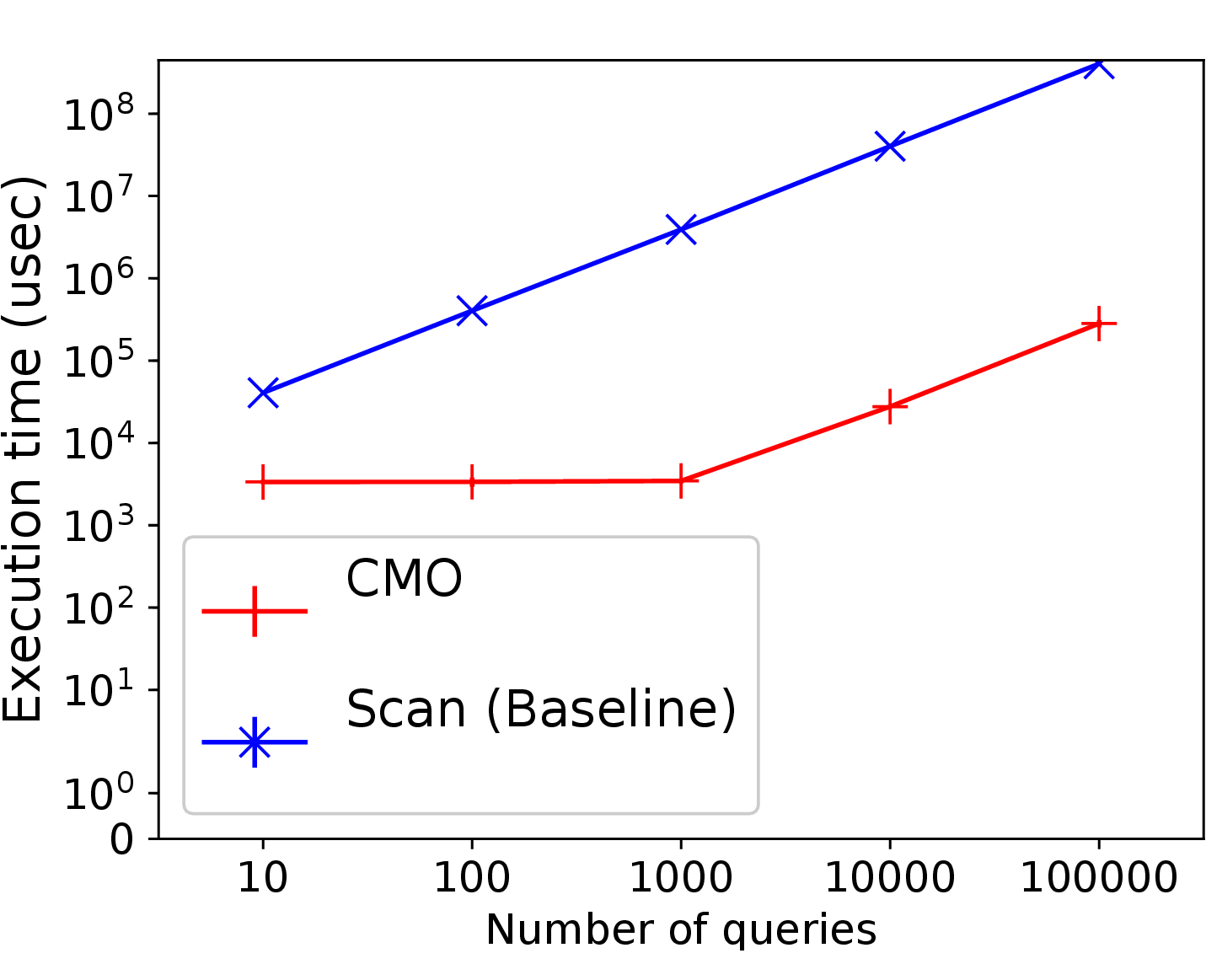}
  \label{exp:bs:query}
}
\subfloat[Binary search: Number of records]{%
  \includegraphics[width=0.25\textwidth]{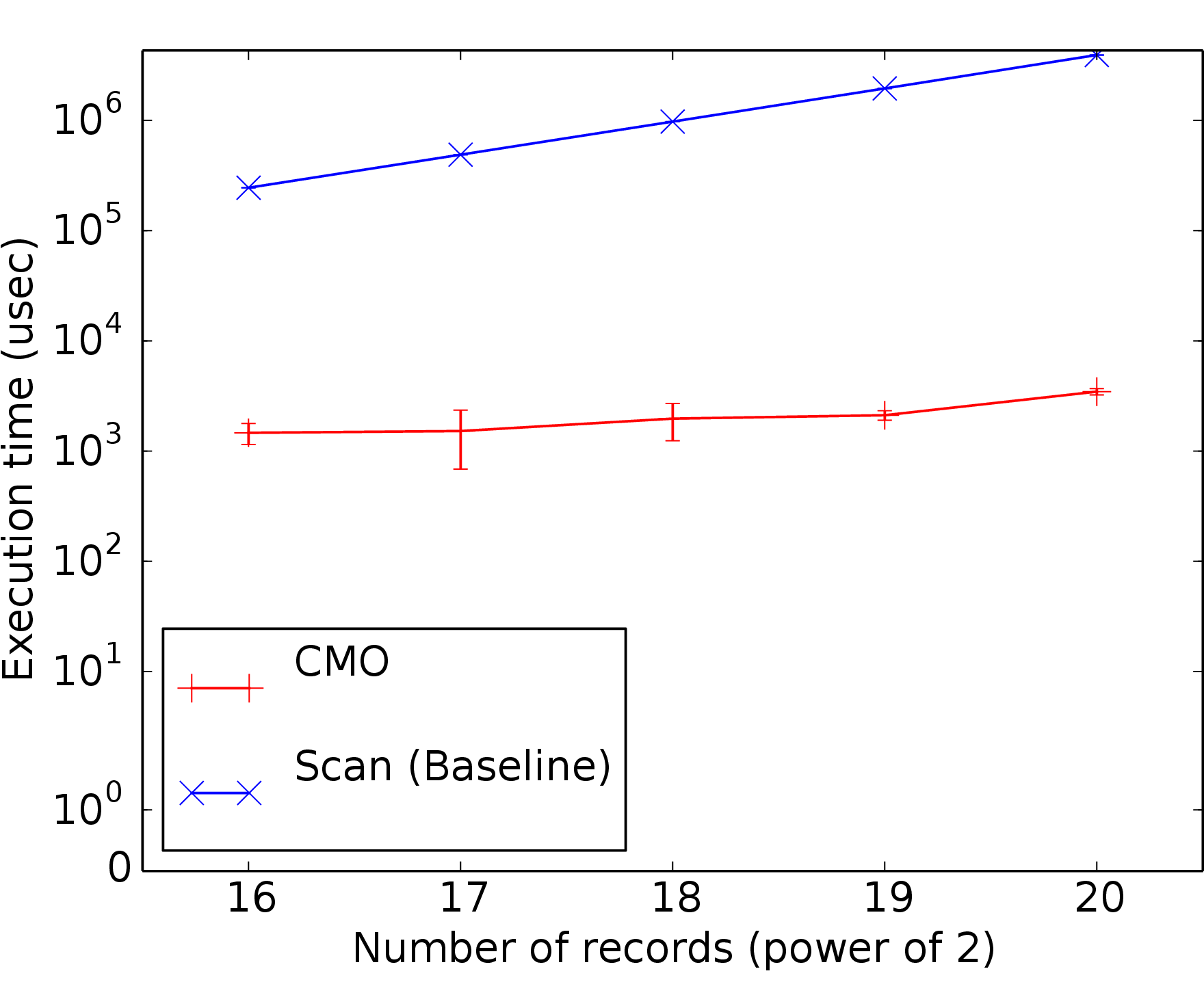}
  \label{exp:bs:data}
}
\subfloat[$K$-means: Number of records]{%
  \includegraphics[width=0.25\textwidth]{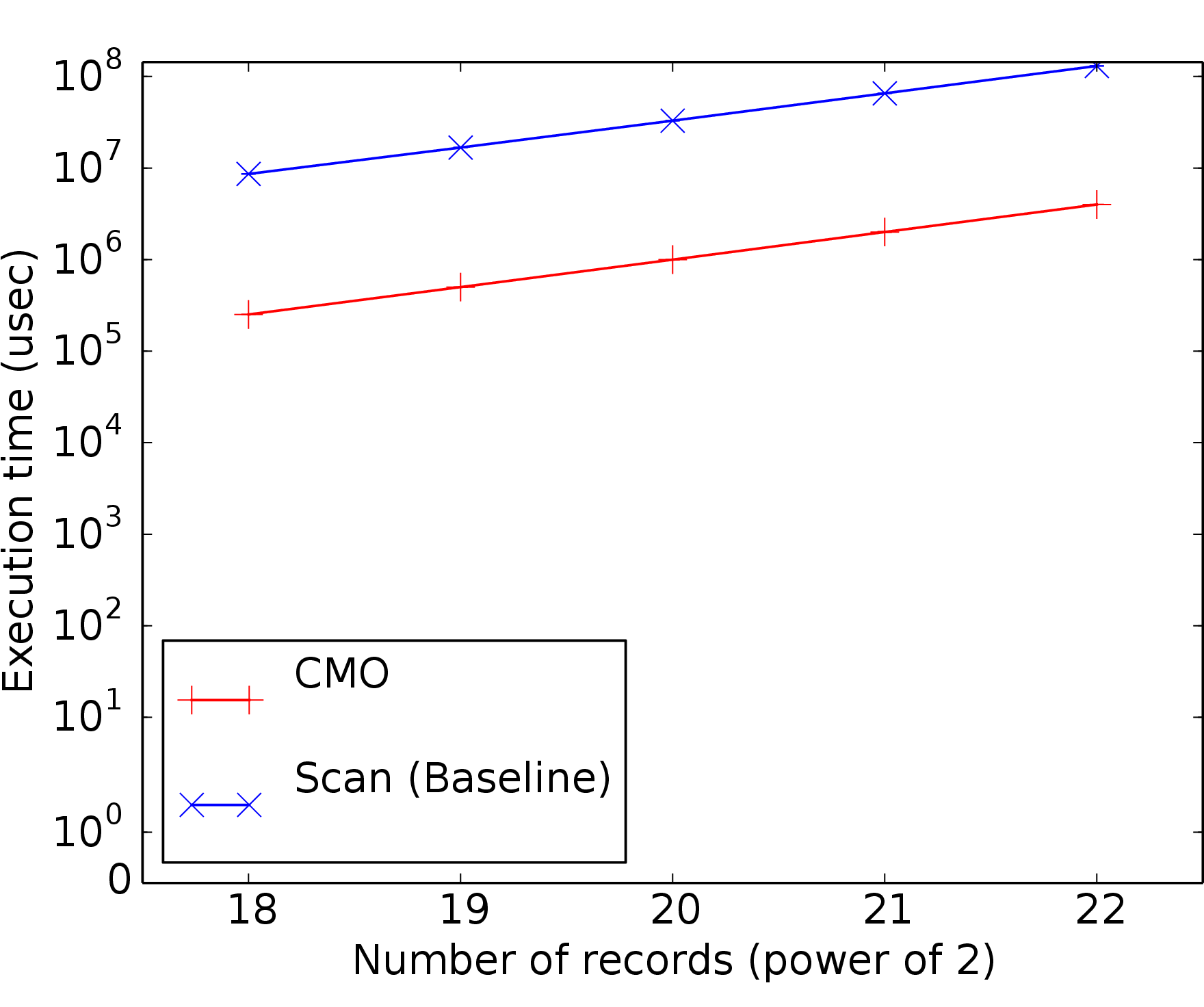}
  \label{exp:kmeans:data}
}
\subfloat[$K$-means: Number of centroids (K)]{%
  \includegraphics[width=0.25\textwidth]{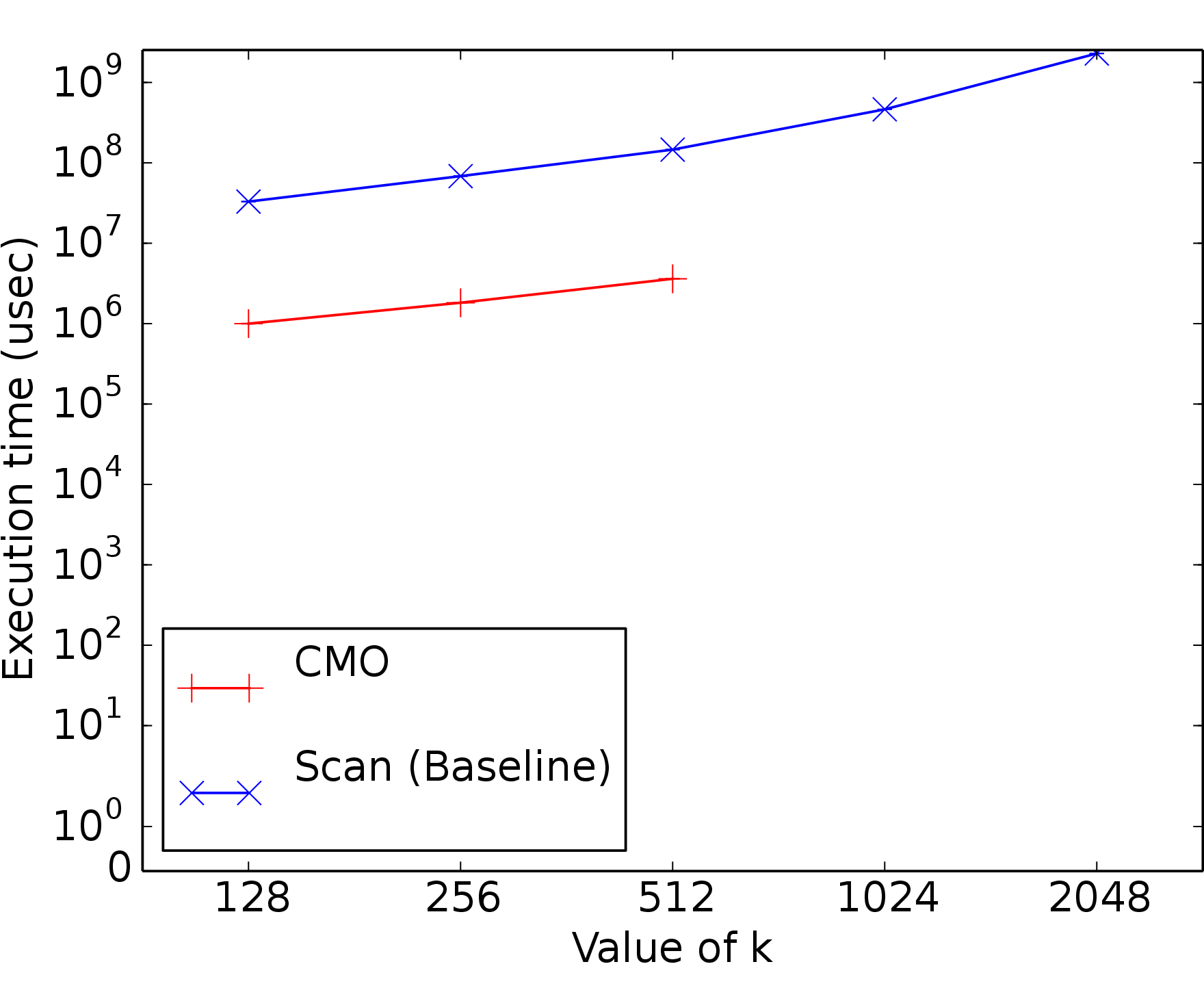}
  \label{exp:kmeans:k}
}
\caption{Comparing CMO with scan: Binary searches and k-means}
\label{exp:vsscan}
\end{figure*}

\noindent{\bf Melbourne shuffle}:
A Melbourne shuffle is a randomized algorithm for data shuffling. A data-shuffle operation takes as input two arrays of equal length, one storing data and the other storing a permutation. It produces the output of permuted data array. For instance, $\texttt{mshuffle}({x,y,z},{1,0,2})={y,x,z}$. Internally, Melbourne shuffle is a randomized algorithm that bucketizes each of the two arrays to $\sqrt{N}$ buckets, each of $\sqrt{N}$ size. A Melbourne shuffle runs in two rounds, each round using $\log{N}\sqrt{N}$ internal memory accessed in a leaky fashion. The original input arrays are in external memory.

To implement Melbourne shuffle using our library, we declare both rounds to be leaky sections. In particular, for scalability, we declare the first round, which is to ``distribute data bucket based on permutation bucket'', to be two leaky sections. Briefly, the first leaky section in the distribution round produces the unordered, linked list as output, without dummy elements. The second leaky section adds dummy elements. In this way, we can improve the data scalability significantly and reduce the access-leaky internal memory from $O(\log{N}\sqrt{N})$ to $O(\sqrt{N})$.   

\noindent{\bf Oblivious merge sort}:
In oblivious merge sort, the operation of merging two sorted lists is made oblivious by a randomized algorithm~\cite{DBLP:conf/ndss/WilliamsS08}. Given two sorted lists, an oblivious merge keeps a $O(\sqrt{N})$ internal-memory buffer and finishes the merge computation with the same $O(N)$ complexity. It is a randomized algorithm and the probability of internal buffer overflowing is made negligible.  

We implement oblivious merge sort by declaring the merge operation to be a leaky section and the internal buffer to be a \texttt{NobRW} and the two merging arrays to be two \texttt{ObRO}.

\noindent{\bf 
Streaming binary search}:
Consider a query on a sorted data array. A binary search locates the matching element in the array to the query. Given a stream of queries, the binary search can be treated as an external-oblivious process in the sense that the internal memory maintains the data array, and external memory maintains the query stream.

To implement the streaming binary search using our library, we declare the binary search to be the leaky section, the data array to be a \texttt{NobRO}, the query stream (array) to be an \texttt{ObRO}. 

\section{Performance Evaluation}

This section evaluates the performance of our system under different sizes of datasets (data scalability). 

%\subsection{Macro Benchmark}

\noindent{\bf Comparison with scan baseline}: 
This set of experiments evaluates the performance of CMO, in comparison to scan-based baselines.  The scan-based baseline translates each leaky random access (on \texttt{NobXX}) to a full array scan. Note that while ORAM~\cite{DBLP:conf/ccs/StefanovDSFRYD13,DBLP:journals/jacm/GoldreichO96} presents a generic solution with better (asymptotic) efficiency, the scan-based approach is state-of-the-art and is more frequently chosen for in-memory data processing, such as oblivious machine learning~\cite{DBLP:conf/uss/OhrimenkoSFMNVC16} and ZeroTrace~\cite{DBLP:journals/iacr/SasyGF17}\footnote{ZeroTrace uses ORAM for disk data access and scan for its internal-memory data access.}. In addition, at a medium data scale, the better asymptotic efficiency of ORAM may not translate to better concrete performance. We implement the scan baseline in our library by converting each access to \texttt{NobXX} to a full-array scan.

We consider the various data analytical computations in the experiments. In each experiment, we measure the performance by execution time. Specifically, the execution time only includes the time spent on running the computation and excludes the time spent on initial data loading (to the cache). We use numeric datasets and generate them randomly. 

We did all the experiments on a laptop with an Intel 8-core i7-6820HK CPU of 2.70GHz, 32KB L1 and 8MB LL cache, 32 GB RAM, and 1 TB Disk. This is one of the Skylake CPUs equipped with both SGX and TSX features. 

\begin{figure}
\subfloat[Melbourne shuffle]{%
  \includegraphics[width=0.25\textwidth]{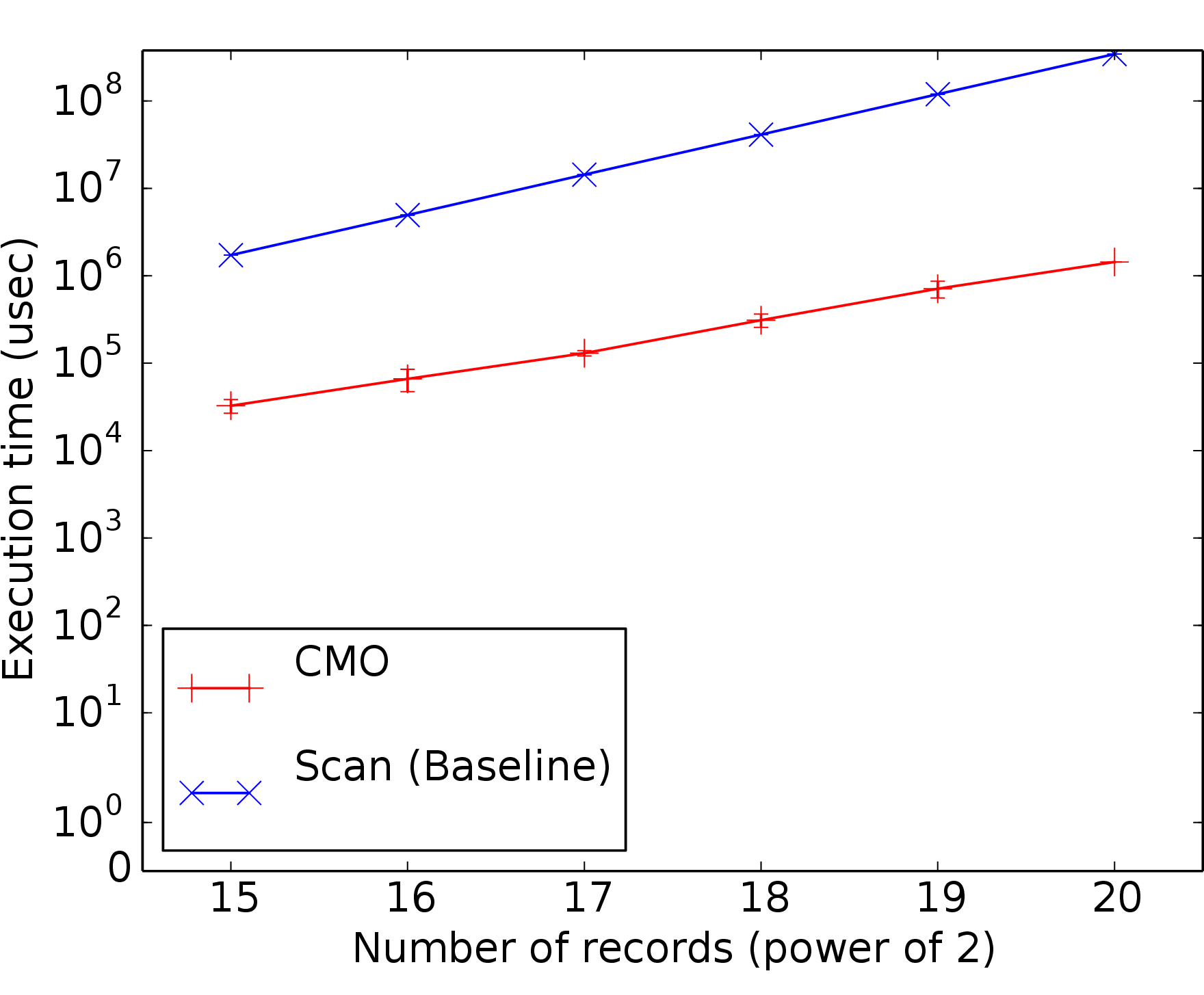}
  \label{exp:melbourne:data}
}
\subfloat[Oblivious merge sort]{%
  \includegraphics[width=0.25\textwidth]{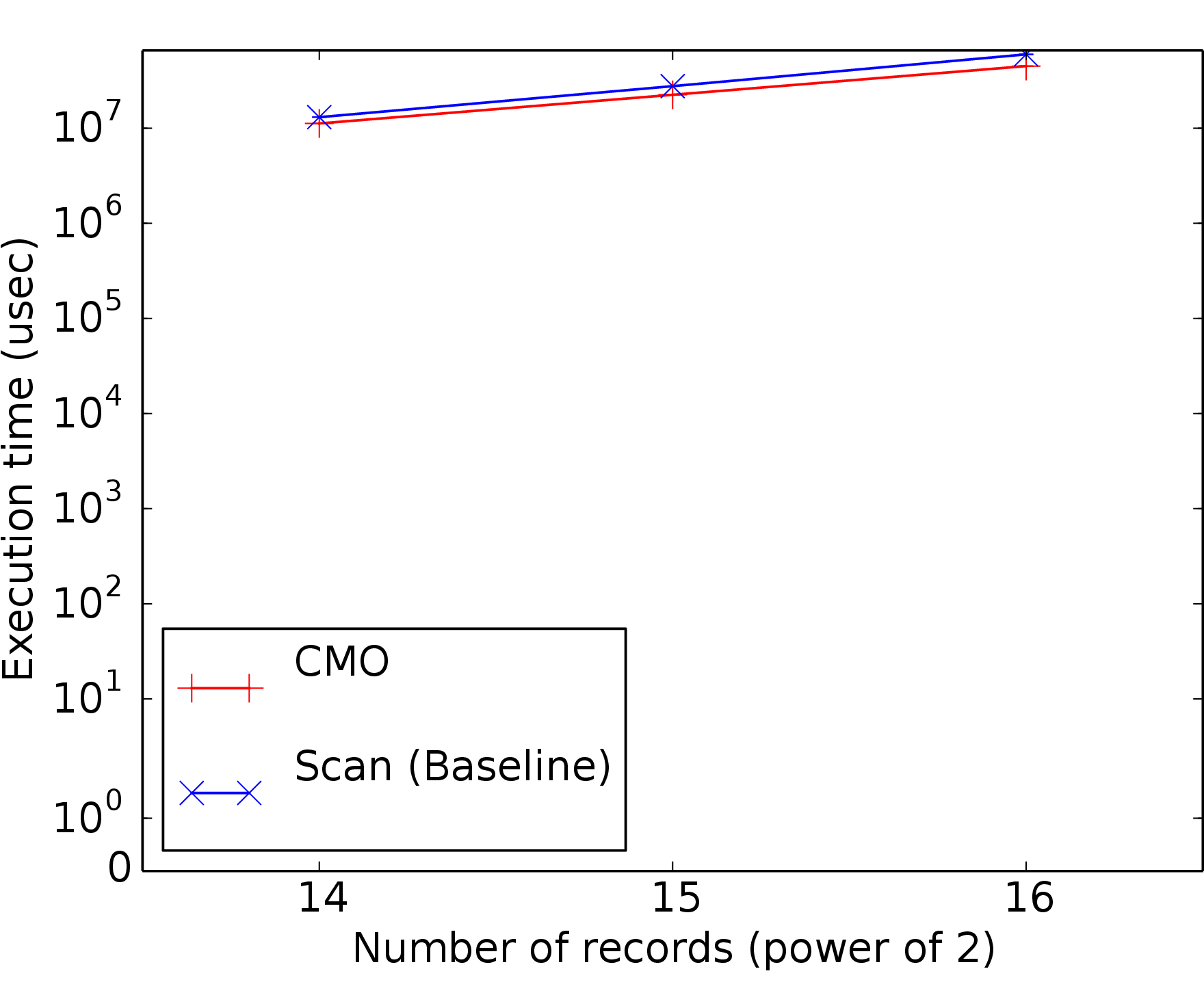}
  \label{exp:oms:data}
}
\caption{Comparing CMO with scan: External-oblivious sorts}
\label{exp:vsscan2}
\end{figure}

\begin{figure}
\centering
\includegraphics[width=0.5\textwidth]{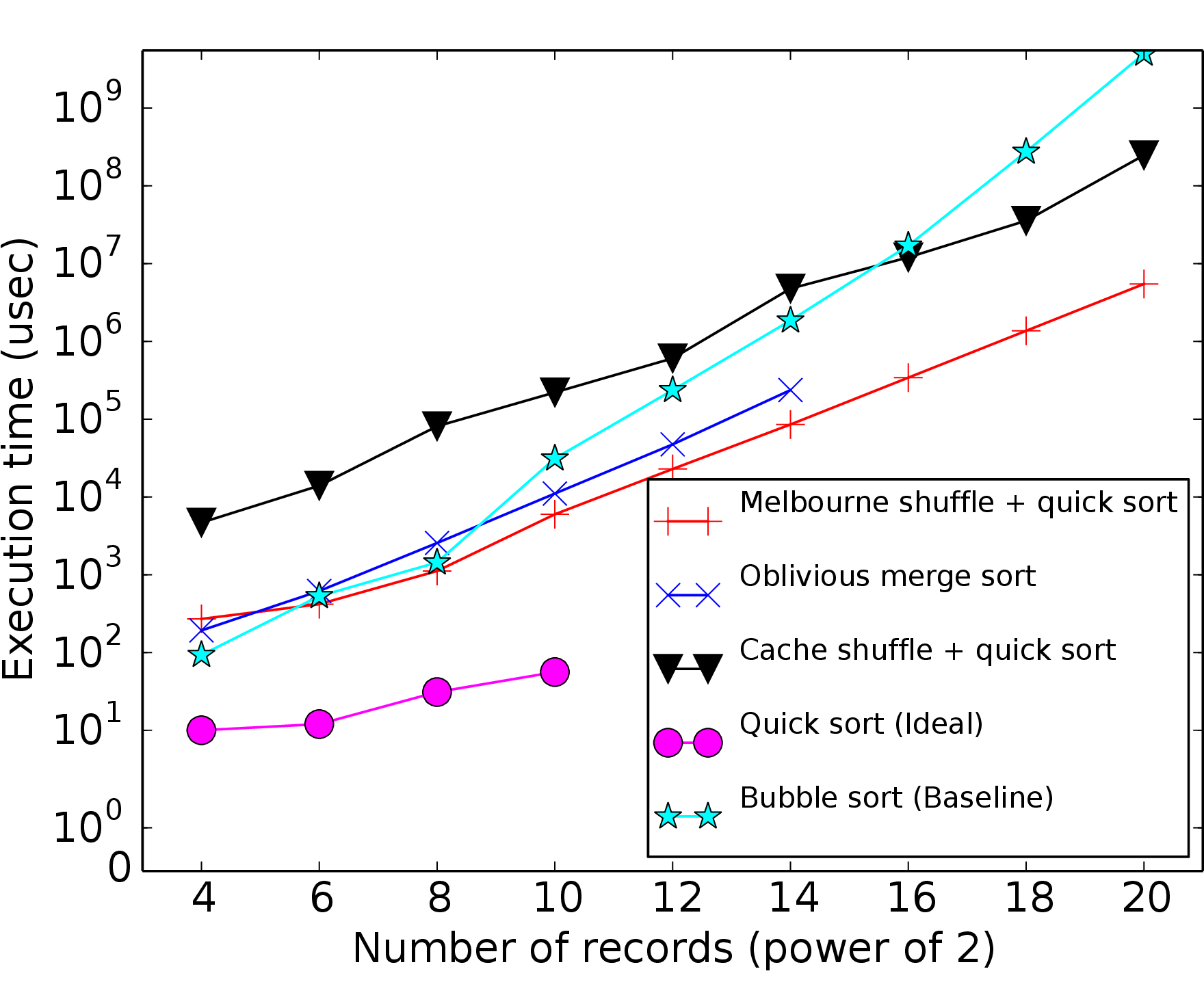}
\caption{Comparing CMO with word-obliviousness and Cloak: the case of sorting}
\label{exp:sorting}
\end{figure}

We vary the data size in each computation and report the results in Figures~\ref{exp:vsscan} and ~\ref{exp:vsscan2}. 
Figure~\ref{exp:melbourne:data} and ~\ref{exp:oms:data} evaluate the execution time of Melbourne shuffle and oblivious merge sort. Compared with the baseline of scan approach, the Melbourne shuffle by CMO achieves a speedup of more than $100\times$ at the largest data scale of 1 million records. For oblivious merge sort, the speedup is much less significant, about $10\%$ speedup of the baseline. 
%The reduced speedup is related to the fact that XXX. 

Figures~\ref{exp:bs:data} and~\ref{exp:bs:query} present the performance result of streaming binary search with varying data size and the number of queries. With varying data size, the CMO outperforms the scan baseline by a speedup ranging from $100\times$ to $1000\times$. With varying query numbers, CMO first keeps the execution time constant until the cache cannot accommodate all the queries. CMO's execution time then linearly scales with the query number. With a large query number, CMO outperforms the scan baseline by $1000\times$ times. Similarly, the performance result of $K$-means in Figures~\ref{exp:kmeans:data} and~\ref{exp:kmeans:k} also show multi-magnitude speedup of CMO.

\noindent{\bf Comparison with other algorithms}:
This set of experiments is to evaluate the performance of CMO against two baselines: 1) word-oblivious algorithms without any TSX transactions, and 2) TSX transactions with no automatic partitioning (e.g., in Cloak~\cite{DBLP:conf/uss/GrussLSOHC17}). For fair comparison, we consider the computation task of sorting and require all approaches achieve the same security level, that is, cache-miss obliviousness. Under this setting, different approaches may run different sorting algorithms. For instance, for 1), the chosen algorithms have to be word oblivious, and we use bubble sort in this baseline. For 2), the algorithms do not need to be oblivious (as the side-channels are defended against by TSX transactions). Thus, we choose the more efficient algorithms, the quick sort. For our work, we use three external oblivious sorting algorithms: Melbourne shuffle with quick sort (following the scramble-then-compute paradigm~\cite{DBLP:journals/popets/DangDCO17}), oblivious merge sort, and cache shuffle with quick-sort. 

The performance result is presented in Figure~\ref{exp:sorting}. The ideal baseline of quick sort running with TSX transactions can only scale to 1024 data records. Because this approach does not partition the program and their computations' working set is limited to the size of CPU cache.
Our CMO approach running oblivious merge sort is limited to support data size of $2^{14}$ records due to the large ``stash'' in the oblivious merge sort algorithm. The other approaches including CMO running Melbourne and cache shuffles, as well as the baseline of word-oblivious bubble sort can scale to large datasets.
In terms of execution time, the CMO approach achieves a speedup by up to two orders of magnitude compared with the word-oblivious bubble sorts (without TSX transactions). This speedup is especially obvious when the data size is more than $2^{14}$ records.

\section{Related Work}

\subsection{Memory-Access Side-Channel Attacks}
\label{sec:background}
\noindent
{\bf Memory-access attacks}: A passive memory-access attacker monitors the access trace of an enclave execution and infers sensitive information. To monitor, the attacker {\it breaks} enclave execution and {\it observes} some micro-architectural side-channel. 
Existing memory-access attacks can be characterized by the spatial resolution of the leakage channel being observed and the temporal resolution of breaking the enclave control flow.
For instance, in the seminal work~\cite{DBLP:conf/sp/XuCP15}, 
a controlled page-fault attacker observes and breaks the enclave execution through page-fault interrupts, where the trace' spatial granularity is page numbers and the temporal granularity is individual page accesses. 
More advanced attacks exploit other leakage channels (e.g., page-table access/dirty bits~\cite{DBLP:conf/uss/BulckWKPS17,DBLP:conf/ccs/WangCPZWBTG17}, shared last-level caches~\cite{DBLP:conf/usenix/HahnelCP17}) to improve spatial and temporal resolution. 
To make it harder to defend against, recent {\it stealthy} attacks break the enclave execution invisibly through hyper-threading that
eliminates interrupts~\cite{DBLP:conf/uss/GrussLSOHC17,DBLP:conf/ccs/WangCPZWBTG17}. 

\ignore{
\begin{figure}
\begin{center}
  \includegraphics[width=0.35\textwidth]{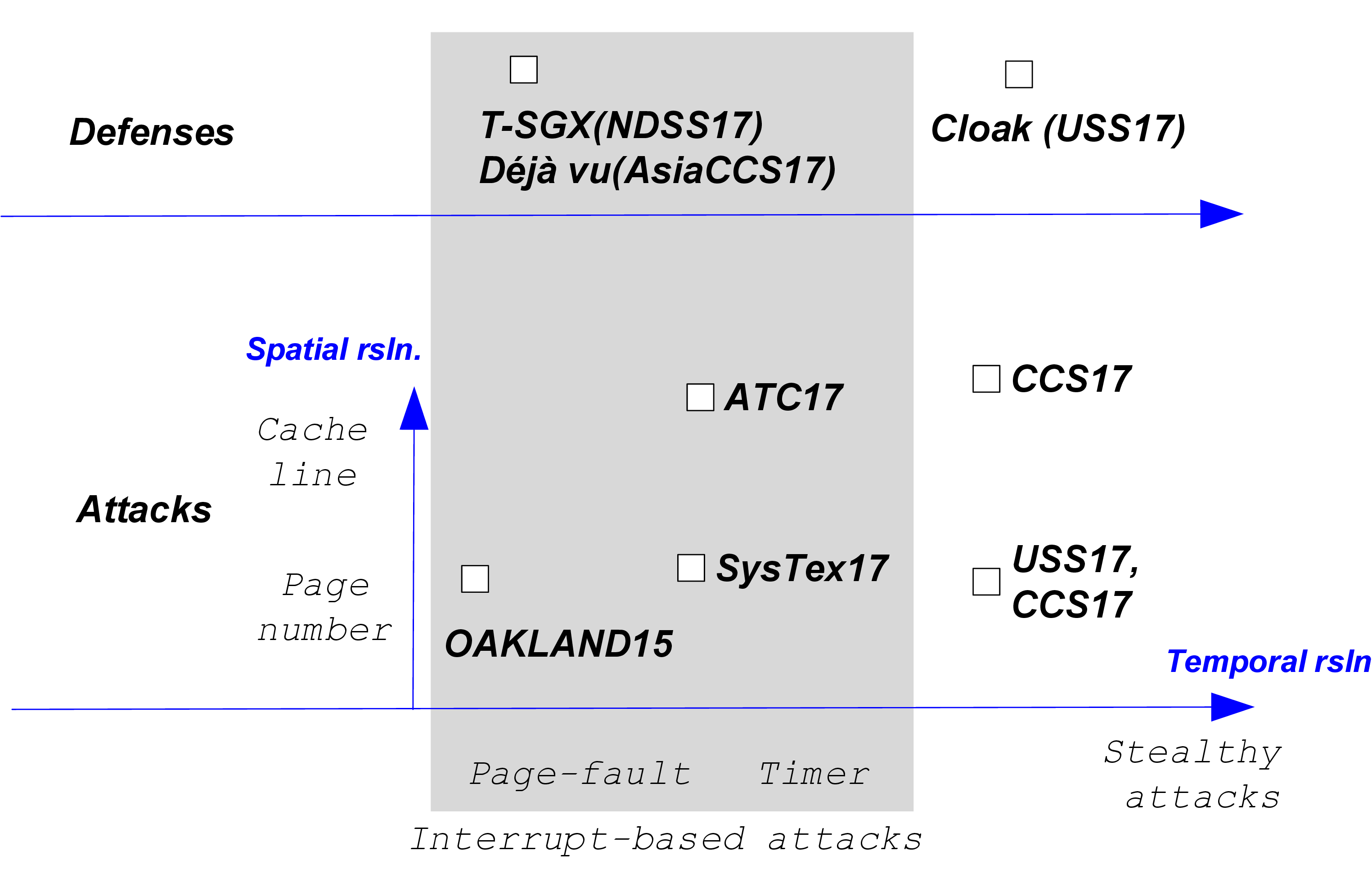}
\caption{Taxonomy of memory-access attacks and defenses in SGX: {\small
Existing attacks shown in the diagram are code-named by the publication venue and year, including USS17 (USENIX Security)~\cite{DBLP:conf/uss/BulckWKPS17}, CCS17~\cite{DBLP:conf/ccs/WangCPZWBTG17}, SysTex17~\cite{DBLP:conf/sosp/BulckPS17}, ATC17~\cite{DBLP:conf/usenix/HahnelCP17}, OAKLAND15~\cite{DBLP:conf/sp/XuCP15}. Existing defenses shown in the diagram are code-named similarly and with their name, including T-SGX~\cite{shih2017t}, D\'ej\`a Vu~\cite{DBLP:conf/ccs/ChenZRZ17}, and Cloak~\cite{DBLP:conf/uss/GrussLSOHC17}.
}}
\label{fig:sgxattacks}
\end{center}
\end{figure}
}

\subsection{Attack Mitigation by Isolation}

\noindent
{\bf Memory-access attack mitigation}:
A successful memory-access attack entails two conditions: 1) an observable channel of memory-access leakage, and 2) the capability of relating the leaked memory-access to sensitive information. A memory-access attack can be mitigated by breaking either condition: M1) Detecting that a leakage channel is being observed by {\it enclave fortification}, or M2) Eliminating the source of leakage by making it difficult to correlate the access trace with sensitive information, via {\it obliviousness}. 

{\color{black}
Existing mitigation techniques take different views regarding how side-channel attack works and take action to close the leakage channel (i.e., isolation). SGX-LAPD~\cite{DBLP:conf/raid/FuBQL17} mitigates the page-fault attacks by enforcing that OS uses large pages. This property  of untrusted OS is verified by the enclave running a program to test the time of page fault. 

T-SGX~\cite{shih2017t} closes the page-fault channel by preventing interrupts in enclaves leveraging the TSX capability. Specifically, by wrapping computation into TSX transactions, any interrupt, including page faults that occur inside a transaction, will be captured by the software. T-SGX runs a policy that bounds the number of interrupts/page-faults to prevent page-fault attacks.

Cloak~\cite{DBLP:conf/uss/GrussLSOHC17} and the work from Chen, et al~\cite{DBLP:conf/ccs/ChenLMZLCW18} close the cache side channels by detecting (individual) cache misses using TSX transactions.

Varys~\cite{DBLP:conf/usenix/OleksenkoTKSF18} reduces all existing side-channel attacks to two leakage channels, namely SGX's Asynchronous EXit (AEX) events and concurrent sharing of CPU cores. To close the first channel, it proposes to monitor the enclave AEX and terminate the enclave execution when the AEX frequency exceeds a threshold. It allocates sufficient hyperthreads to keep a CPU core occupied by a single application (so-called thread-core reservation) to close the second channel. Vary enforces the two conditions (AEX monitoring and thread-core reservation) in a trust-but-verify manner, that is, the OS is trusted to execute the program with the required property verified nevertheless by the enclave.
}

For the former, existing enclave-fortification work detects the attack on the events of page-fault~\cite{DBLP:conf/ccs/ShindeCNS16,shih2017t}, cache-miss~\cite{DBLP:conf/uss/GrussLSOHC17,DBLP:conf/woot/BrasserMDKCS17}, etc. 

For the latter, existing work on {\bf data-oblivious systems} is based on oblivious algorithms~\cite{DBLP:conf/nsdi/ZhengDBPGS17,DBLP:conf/ccs/OhrimenkoCFGKS15} or less-efficient oblivious RAM~\cite{DBLP:journals/iacr/SasyGF17,2017arXiv171000458E}.
Concretely, data obliviousness can eliminate the source of leakage in the presence of memory-access attacks on TEE. When building data-oblivious systems, a design choice is which oblivious-computing mechanisms to use: Opaque~\cite{DBLP:conf/nsdi/ZhengDBPGS17} supports database queries by engineering word-oblivious algorithms (Column Sort~\cite{DBLP:journals/siamcomp/AzarV87}). Oblivious machine-learning~\cite{DBLP:conf/uss/OhrimenkoSFMNVC16} combines word-oblivious algorithms and the trivial mechanism of converting random-access to a full-array scan. Scan-based transformation is similarly adopted in building ZeroTrace~\cite{DBLP:journals/iacr/SasyGF17}, an ORAM system in SGX. ObliVM~\cite{DBLP:conf/sp/LiuWNHS15} is a compiler that supports the engineering of both word-oblivious algorithms and ORAM~\cite{DBLP:conf/ccs/StefanovDSFRYD13,DBLP:journals/jacm/GoldreichO96}. Oblivious Map-Reduce~\cite{DBLP:conf/ccs/OhrimenkoCFGKS15} and Prochlo~\cite{DBLP:conf/sosp/BittauEMMRLRKTS17} use external oblivious shuffle~\cite{DBLP:conf/icalp/OhrimenkoGTU14} to protect distributed systems under traffic analysis, but insecure under micro-architectural memory-access attacks. 

The related work~\cite{me:cmos} presents the preliminary idea of dynamic program partitioning but is limited to a case study on a specific computation (data array shuffling). By comparison, this work presents a tool for generic computation.

\subsection{Attack Mitigation by Data Obliviousness}

\noindent
{\bf Data-oblivious algorithms}:  The conventional oblivious algorithms ensure oblivious data access at the ``word'' granularity. These {\it word-oblivious algorithms (A1)} are constructed based on primitives such as compare-exchanges and tend to be not very practical (e.g., AKS sorting~\cite{DBLP:conf/stoc/AjtaiKS83} and sub-optimal column sort~\cite{DBLP:journals/siamcomp/AzarV87}). In the external-memory model, {\bf external oblivious algorithms} improve the time/IO complexity by assuming a small amount of internal storage that is accessed in a leaky fashion. Due to a smaller constant and optimal complexity, this family of algorithms is of great practical interest. For instance,  Melbourne shuffle~\cite{DBLP:conf/icalp/OhrimenkoGTU14} causes $O(\sqrt{N})$ IO with $O(N\log{N})$ time at the expense of $\sqrt{N}$ internal space and was recently used in constructing practical systems (e.g., MapReduce~\cite{DBLP:conf/ccs/OhrimenkoCFGKS15}). The {\it oblivious RAM (A3)}~\cite{DBLP:conf/ccs/StefanovDSFRYD13,DBLP:journals/jacm/GoldreichO96} is an external data structure supporting oblivious external data access by translating virtual random-access to oblivious physical data access with poly-logarithmic blowup.

Compared with ORAM (A3) and word-oblivious algorithms (A1), the external oblivious algorithms (A2) cause lower overhead in complexity. Compared with generic ORAM, external oblivious algorithms are computation specific; existing algorithms support sort, compaction, and relational queries and other batched-oriented computations~\cite{DBLP:conf/icalp/OhrimenkoGTU14,DBLP:conf/ndss/WilliamsS08,DBLP:conf/spaa/Goodrich11,DBLP:conf/icdt/ArasuK14}.

\ignore{
\begin{table*}[!htbp] 
\caption{Research distinction}
\label{tab:distinct}\centering{
\small
\begin{tabularx}{0.9\textwidth}{|c|c|X|X|X|}
  \hline
 & Method
 & Cache access-pattern security 
 & Data scalability 
 & Algorithmic complexity \\ \hline
  T-SGX~\cite{shih2017t}, Cloak~\cite{DBLP:conf/uss/GrussLSOHC17}
  & Attack detection by HTM & + & - & + \\ \hline
  Opaque~\cite{DBLP:conf/nsdi/ZhengDBPGS17}, ObliviousML~\cite{DBLP:conf/uss/OhrimenkoSFMNVC16}
  & Word-oblivious algo. & + & + & - \\ \hline
  ObliviousMR~\cite{DBLP:conf/ccs/OhrimenkoCFGKS15},Prochlo~\cite{DBLP:conf/sosp/BittauEMMRLRKTS17}
  & External oblivious algo. & - & + & + \\ \hline
  ZeroTrace~\cite{DBLP:journals/iacr/SasyGF17}
  & External ORAM & + & + & - \\ \hline
  This work
  & External oblivious algo./HTM & + & + & + \\ \hline
\end{tabularx}
}
\end{table*}
}

\section{Conclusion}
This work enables cache-miss oblivious data analytics with Intel SGX. It considers expressing target data analytics by external oblivious algorithms and protects their stash's memory access pattern by dynamically partitioning the code and wrapping them in TSX transactions. 
We develop a C++ library that monitors  memory accesses and builds a cache model to predict cache misses. 
On the library, we implement various external oblivious algorithms.
The evaluation shows effectively increased transaction size and reduced execution time by orders of magnitude in comparison with the state-of-the-art systems.

%\input{task11_full/TOREMOVE/boa_v2.tex}

%\input{text/wheels.tex}

\bibliographystyle{ACM-Reference-Format}
\bibliography{ads,lsm,yuzhetang,bkc,odb,cacheattacks,sc,crypto,sgx,diffpriv,txtbk,distrkvs,vc}


\begin{thebibliography}{37}

%%% the Web address.

\ifx \showCODEN    \undefined \def \showCODEN     #1{\unskip}     \fi
\ifx \showDOI      \undefined \def \showDOI       #1{#1}\fi
\ifx \showISBNx    \undefined \def \showISBNx     #1{\unskip}     \fi
\ifx \showISBNxiii \undefined \def \showISBNxiii  #1{\unskip}     \fi
\ifx \showISSN     \undefined \def \showISSN      #1{\unskip}     \fi
\ifx \showLCCN     \undefined \def \showLCCN      #1{\unskip}     \fi
\ifx \shownote     \undefined \def \shownote      #1{#1}          \fi
\ifx \showarticletitle \undefined \def \showarticletitle #1{#1}   \fi
\ifx \showURL      \undefined \def \showURL       {\relax}        \fi
% invisible to TeX
\providecommand\bibfield[2]{#2}
\providecommand\bibinfo[2]{#2}
\providecommand\natexlab[1]{#1}
\providecommand\showeprint[2][]{arXiv:#2}

\bibitem[\protect\citeauthoryear{??}{me:}{[n.d.]a}]%
        {me:tzone}
 \bibinfo{year}{[n.d.]}\natexlab{a}.
\newblock \showarticletitle{{ARM TrustZone},
  https://www.arm.com/products/security-on-arm/trustzone}.
\newblock


\bibitem[\protect\citeauthoryear{??}{me:}{[n.d.]b}]%
        {me:scpu}
 \bibinfo{year}{[n.d.]}\natexlab{b}.
\newblock \showarticletitle{{IBM SCPU},
  http://www-03.ibm.com/security/cryptocards/}.
\newblock


\bibitem[\protect\citeauthoryear{??}{me:}{[n.d.]c}]%
        {me:txt}
 \bibinfo{year}{[n.d.]}\natexlab{c}.
\newblock \showarticletitle{{Intel TXT},
  http://www.intel.com/technology/security/
  downloads/TrustedExec\_Overview.pdf}.
\newblock


\bibitem[\protect\citeauthoryear{??}{me:}{[n.d.]d}]%
        {me:tpm}
 \bibinfo{year}{[n.d.]}\natexlab{d}.
\newblock \showarticletitle{{TPM},
  http://www.trustedcomputinggroup.org/tpm-main-specification/}.
\newblock


\bibitem[\protect\citeauthoryear{??}{DBL}{2015}]%
        {DBLP:conf/sp/2015}
 \bibinfo{year}{2015}\natexlab{}.
\newblock \bibinfo{booktitle}{\emph{2015 {IEEE} Symposium on Security and
  Privacy, {SP} 2015, San Jose, CA, USA, May 17-21, 2015}}.
  \bibinfo{publisher}{{IEEE} Computer Society}.
\newblock
\showISBNx{978-1-4673-6949-7}
\urldef\tempurl%
\url{http://ieeexplore.ieee.org/xpl/mostRecentIssue.jsp?punumber=7160813}
\showURL{%
\tempurl}


\bibitem[\protect\citeauthoryear{Ajtai, Koml{\'{o}}s, and
  Szemer{\'{e}}di}{Ajtai et~al\mbox{.}}{1983}]%
        {DBLP:conf/stoc/AjtaiKS83}
\bibfield{author}{\bibinfo{person}{Mikl{\'{o}}s Ajtai},
  \bibinfo{person}{J{\'{a}}nos Koml{\'{o}}s}, {and} \bibinfo{person}{Endre
  Szemer{\'{e}}di}.} \bibinfo{year}{1983}\natexlab{}.
\newblock \showarticletitle{An O(n log n) Sorting Network}. In
  \bibinfo{booktitle}{\emph{Proceedings of the 15th Annual {ACM} Symposium on
  Theory of Computing, 25-27 April, 1983, Boston, Massachusetts, {USA}}},
  \bibfield{editor}{\bibinfo{person}{David~S. Johnson}, \bibinfo{person}{Ronald
  Fagin}, \bibinfo{person}{Michael~L. Fredman}, \bibinfo{person}{David Harel},
  \bibinfo{person}{Richard~M. Karp}, \bibinfo{person}{Nancy~A. Lynch},
  \bibinfo{person}{Christos~H. Papadimitriou}, \bibinfo{person}{Ronald~L.
  Rivest}, \bibinfo{person}{Walter~L. Ruzzo}, {and} \bibinfo{person}{Joel~I.
  Seiferas}} (Eds.). \bibinfo{publisher}{{ACM}}, \bibinfo{pages}{1--9}.
\newblock
\urldef\tempurl%
\url{https://doi.org/10.1145/800061.808726}
\showDOI{\tempurl}


\bibitem[\protect\citeauthoryear{Anati, Gueron, Johnson, and Scarlata}{Anati
  et~al\mbox{.}}{[n.d.]}]%
        {Anati_innovativetechnology}
\bibfield{author}{\bibinfo{person}{Ittai Anati}, \bibinfo{person}{Shay Gueron},
  \bibinfo{person}{Simon~P Johnson}, {and} \bibinfo{person}{Vincent~R
  Scarlata}.} \bibinfo{year}{[n.d.]}\natexlab{}.
\newblock \bibinfo{title}{Innovative Technology for CPU Based Attestation and
  Sealing}.
\newblock
\newblock


\bibitem[\protect\citeauthoryear{Arasu and Kaushik}{Arasu and Kaushik}{2014}]%
        {DBLP:conf/icdt/ArasuK14}
\bibfield{author}{\bibinfo{person}{Arvind Arasu} {and} \bibinfo{person}{Raghav
  Kaushik}.} \bibinfo{year}{2014}\natexlab{}.
\newblock \showarticletitle{Oblivious Query Processing}. In
  \bibinfo{booktitle}{\emph{Proc. 17th International Conference on Database
  Theory (ICDT), Athens, Greece, March 24-28, 2014.}} \bibinfo{pages}{26--37}.
\newblock
\urldef\tempurl%
\url{https://doi.org/10.5441/002/icdt.2014.07}
\showDOI{\tempurl}


\bibitem[\protect\citeauthoryear{Azar and Vishkin}{Azar and Vishkin}{1987}]%
        {DBLP:journals/siamcomp/AzarV87}
\bibfield{author}{\bibinfo{person}{Yossi Azar} {and} \bibinfo{person}{Uzi
  Vishkin}.} \bibinfo{year}{1987}\natexlab{}.
\newblock \showarticletitle{Tight Comparison Bounds on the Complexity of
  Parallel Sorting}.
\newblock \bibinfo{journal}{\emph{{SIAM} J. Comput.}} \bibinfo{volume}{16},
  \bibinfo{number}{3} (\bibinfo{year}{1987}), \bibinfo{pages}{458--464}.
\newblock
\urldef\tempurl%
\url{https://doi.org/10.1137/0216032}
\showDOI{\tempurl}


\bibitem[\protect\citeauthoryear{Bittau, Erlingsson, Maniatis, Mironov,
  Raghunathan, Lie, Rudominer, Kode, Tinn{\'{e}}s, and Seefeld}{Bittau
  et~al\mbox{.}}{2017}]%
        {DBLP:conf/sosp/BittauEMMRLRKTS17}
\bibfield{author}{\bibinfo{person}{Andrea Bittau}, \bibinfo{person}{{\'{U}}lfar
  Erlingsson}, \bibinfo{person}{Petros Maniatis}, \bibinfo{person}{Ilya
  Mironov}, \bibinfo{person}{Ananth Raghunathan}, \bibinfo{person}{David Lie},
  \bibinfo{person}{Mitch Rudominer}, \bibinfo{person}{Ushasree Kode},
  \bibinfo{person}{Julien Tinn{\'{e}}s}, {and} \bibinfo{person}{Bernhard
  Seefeld}.} \bibinfo{year}{2017}\natexlab{}.
\newblock \showarticletitle{Prochlo: Strong Privacy for Analytics in the
  Crowd}. In \bibinfo{booktitle}{\emph{Proceedings of the 26th Symposium on
  Operating Systems Principles, Shanghai, China, October 28-31, 2017}}.
  \bibinfo{publisher}{{ACM}}, \bibinfo{pages}{441--459}.
\newblock
\showISBNx{978-1-4503-5085-3}
\urldef\tempurl%
\url{https://doi.org/10.1145/3132747.3132769}
\showDOI{\tempurl}


\bibitem[\protect\citeauthoryear{Brasser, M{\"{u}}ller, Dmitrienko, Kostiainen,
  Capkun, and Sadeghi}{Brasser et~al\mbox{.}}{2017}]%
        {DBLP:conf/woot/BrasserMDKCS17}
\bibfield{author}{\bibinfo{person}{Ferdinand Brasser}, \bibinfo{person}{Urs
  M{\"{u}}ller}, \bibinfo{person}{Alexandra Dmitrienko}, \bibinfo{person}{Kari
  Kostiainen}, \bibinfo{person}{Srdjan Capkun}, {and}
  \bibinfo{person}{Ahmad{-}Reza Sadeghi}.} \bibinfo{year}{2017}\natexlab{}.
\newblock \showarticletitle{Software Grand Exposure: {SGX} Cache Attacks Are
  Practical}. In \bibinfo{booktitle}{\emph{11th {USENIX} Workshop on Offensive
  Technologies, {WOOT} 2017, Vancouver, BC, Canada, August 14-15, 2017.}},
  \bibfield{editor}{\bibinfo{person}{William Enck} {and}
  \bibinfo{person}{Collin Mulliner}} (Eds.). \bibinfo{publisher}{{USENIX}
  Association}.
\newblock
\urldef\tempurl%
\url{https://www.usenix.org/conference/woot17/workshop-program/presentation/brasser}
\showURL{%
\tempurl}


\bibitem[\protect\citeauthoryear{Chen, Tang, and Zhou}{Chen
  et~al\mbox{.}}{2017a}]%
        {me:cmos}
\bibfield{author}{\bibinfo{person}{Ju Chen}, \bibinfo{person}{Yuzhe~(Richard)
  Tang}, {and} \bibinfo{person}{Hao Zhou}.} \bibinfo{year}{2017}\natexlab{a}.
\newblock \showarticletitle{Strongly Secure and Efficient Data Shuffle on
  Hardware Enclaves}.
\newblock \bibinfo{journal}{\emph{ACM SOSP Workshop (SysTex)}}
  (\bibinfo{year}{2017}).
\newblock


\bibitem[\protect\citeauthoryear{Chen, Liu, Mi, Zhang, Lee, Chen, and
  Wang}{Chen et~al\mbox{.}}{2018}]%
        {DBLP:conf/ccs/ChenLMZLCW18}
\bibfield{author}{\bibinfo{person}{Sanchuan Chen}, \bibinfo{person}{Fangfei
  Liu}, \bibinfo{person}{Zeyu Mi}, \bibinfo{person}{Yinqian Zhang},
  \bibinfo{person}{Ruby~B. Lee}, \bibinfo{person}{Haibo Chen}, {and}
  \bibinfo{person}{XiaoFeng Wang}.} \bibinfo{year}{2018}\natexlab{}.
\newblock \showarticletitle{Leveraging Hardware Transactional Memory for Cache
  Side-Channel Defenses}. In \bibinfo{booktitle}{\emph{Proceedings of the 2018
  on Asia Conference on Computer and Communications Security, AsiaCCS 2018,
  Incheon, Republic of Korea, June 04-08, 2018}},
  \bibfield{editor}{\bibinfo{person}{Jong Kim}, \bibinfo{person}{Gail{-}Joon
  Ahn}, \bibinfo{person}{Seungjoo Kim}, \bibinfo{person}{Yongdae Kim},
  \bibinfo{person}{Javier L{\'{o}}pez}, {and} \bibinfo{person}{Taesoo Kim}}
  (Eds.). \bibinfo{publisher}{{ACM}}, \bibinfo{pages}{601--608}.
\newblock
\urldef\tempurl%
\url{https://doi.org/10.1145/3196494.3196501}
\showDOI{\tempurl}


\bibitem[\protect\citeauthoryear{Chen, Zhang, Reiter, and Zhang}{Chen
  et~al\mbox{.}}{2017b}]%
        {DBLP:conf/ccs/ChenZRZ17}
\bibfield{author}{\bibinfo{person}{Sanchuan Chen}, \bibinfo{person}{Xiaokuan
  Zhang}, \bibinfo{person}{Michael~K. Reiter}, {and} \bibinfo{person}{Yinqian
  Zhang}.} \bibinfo{year}{2017}\natexlab{b}.
\newblock \showarticletitle{Detecting Privileged Side-Channel Attacks in
  Shielded Execution with D{\'{e}}j{\`{a}} Vu}. In
  \bibinfo{booktitle}{\emph{Proceedings of the 2017 {ACM} on Asia Conference on
  Computer and Communications Security, AsiaCCS 2017, Abu Dhabi, United Arab
  Emirates, April 2-6, 2017}}, \bibfield{editor}{\bibinfo{person}{Ramesh
  Karri}, \bibinfo{person}{Ozgur Sinanoglu}, \bibinfo{person}{Ahmad{-}Reza
  Sadeghi}, {and} \bibinfo{person}{Xun Yi}} (Eds.). \bibinfo{publisher}{{ACM}},
  \bibinfo{pages}{7--18}.
\newblock
\showISBNx{978-1-4503-4944-4}
\urldef\tempurl%
\url{https://doi.org/10.1145/3052973.3053007}
\showDOI{\tempurl}


\bibitem[\protect\citeauthoryear{Dang, Dinh, Chang, and Ooi}{Dang
  et~al\mbox{.}}{2017}]%
        {DBLP:journals/popets/DangDCO17}
\bibfield{author}{\bibinfo{person}{Hung Dang}, \bibinfo{person}{Tien Tuan~Anh
  Dinh}, \bibinfo{person}{Ee{-}Chien Chang}, {and} \bibinfo{person}{Beng~Chin
  Ooi}.} \bibinfo{year}{2017}\natexlab{}.
\newblock \showarticletitle{Privacy-Preserving Computation with Trusted
  Computing via Scramble-then-Compute}.
\newblock \bibinfo{journal}{\emph{PoPETs}} \bibinfo{volume}{2017},
  \bibinfo{number}{3} (\bibinfo{year}{2017}), \bibinfo{pages}{21}.
\newblock
\urldef\tempurl%
\url{https://doi.org/10.1515/popets-2017-0026}
\showDOI{\tempurl}


\bibitem[\protect\citeauthoryear{{Eskandarian} and {Zaharia}}{{Eskandarian} and
  {Zaharia}}{2017}]%
        {2017arXiv171000458E}
\bibfield{author}{\bibinfo{person}{S. {Eskandarian}} {and} \bibinfo{person}{M.
  {Zaharia}}.} \bibinfo{year}{2017}\natexlab{}.
\newblock \showarticletitle{{An Oblivious General-Purpose SQL Database for the
  Cloud}}.
\newblock \bibinfo{journal}{\emph{ArXiv e-prints}} (\bibinfo{date}{Oct.}
  \bibinfo{year}{2017}).
\newblock
\showeprint[arxiv]{cs.CR/1710.00458}


\bibitem[\protect\citeauthoryear{Fu, Bauman, Quinonez, and Lin}{Fu
  et~al\mbox{.}}{2017}]%
        {DBLP:conf/raid/FuBQL17}
\bibfield{author}{\bibinfo{person}{Yangchun Fu}, \bibinfo{person}{Erick
  Bauman}, \bibinfo{person}{Raul Quinonez}, {and} \bibinfo{person}{Zhiqiang
  Lin}.} \bibinfo{year}{2017}\natexlab{}.
\newblock \showarticletitle{Sgx-Lapd: Thwarting Controlled Side Channel Attacks
  via Enclave Verifiable Page Faults}. In \bibinfo{booktitle}{\emph{Research in
  Attacks, Intrusions, and Defenses - 20th International Symposium, {RAID}
  2017, Atlanta, GA, USA, September 18-20, 2017, Proceedings}}
  \emph{(\bibinfo{series}{Lecture Notes in Computer Science})},
  \bibfield{editor}{\bibinfo{person}{Marc Dacier}, \bibinfo{person}{Michael
  Bailey}, \bibinfo{person}{Michalis Polychronakis}, {and}
  \bibinfo{person}{Manos Antonakakis}} (Eds.), Vol.~\bibinfo{volume}{10453}.
  \bibinfo{publisher}{Springer}, \bibinfo{pages}{357--380}.
\newblock
\showISBNx{978-3-319-66331-9}
\urldef\tempurl%
\url{https://doi.org/10.1007/978-3-319-66332-6\_16}
\showDOI{\tempurl}


\bibitem[\protect\citeauthoryear{Goldreich and Ostrovsky}{Goldreich and
  Ostrovsky}{1996}]%
        {DBLP:journals/jacm/GoldreichO96}
\bibfield{author}{\bibinfo{person}{Oded Goldreich} {and}
  \bibinfo{person}{Rafail Ostrovsky}.} \bibinfo{year}{1996}\natexlab{}.
\newblock \showarticletitle{Software Protection and Simulation on Oblivious
  RAMs}.
\newblock \bibinfo{journal}{\emph{J. {ACM}}} \bibinfo{volume}{43},
  \bibinfo{number}{3} (\bibinfo{year}{1996}), \bibinfo{pages}{431--473}.
\newblock
\urldef\tempurl%
\url{https://doi.org/10.1145/233551.233553}
\showDOI{\tempurl}


\bibitem[\protect\citeauthoryear{Goodrich}{Goodrich}{2011}]%
        {DBLP:conf/spaa/Goodrich11}
\bibfield{author}{\bibinfo{person}{Michael~T. Goodrich}.}
  \bibinfo{year}{2011}\natexlab{}.
\newblock \showarticletitle{Data-oblivious external-memory algorithms for the
  compaction, selection, and sorting of outsourced data}. In
  \bibinfo{booktitle}{\emph{{SPAA} 2011: Proceedings of the 23rd Annual {ACM}
  Symposium on Parallelism in Algorithms and Architectures, San Jose, CA, USA,
  June 4-6, 2011 (Co-located with {FCRC} 2011)}}. \bibinfo{pages}{379--388}.
\newblock
\urldef\tempurl%
\url{https://doi.org/10.1145/1989493.1989555}
\showDOI{\tempurl}


\bibitem[\protect\citeauthoryear{Gruss, Lettner, Schuster, Ohrimenko, Haller,
  and Costa}{Gruss et~al\mbox{.}}{2017}]%
        {DBLP:conf/uss/GrussLSOHC17}
\bibfield{author}{\bibinfo{person}{Daniel Gruss}, \bibinfo{person}{Julian
  Lettner}, \bibinfo{person}{Felix Schuster}, \bibinfo{person}{Olga Ohrimenko},
  \bibinfo{person}{Istv{\'{a}}n Haller}, {and} \bibinfo{person}{Manuel Costa}.}
  \bibinfo{year}{2017}\natexlab{}.
\newblock \showarticletitle{Strong and Efficient Cache Side-Channel Protection
  using Hardware Transactional Memory}, See \citeN{DBLP:conf/uss/2017},
  \bibinfo{pages}{217--233}.
\newblock
\urldef\tempurl%
\url{https://www.usenix.org/conference/usenixsecurity17/technical-sessions/presentation/gruss}
\showURL{%
\tempurl}


\bibitem[\protect\citeauthoryear{H{\"{a}}hnel, Cui, and Peinado}{H{\"{a}}hnel
  et~al\mbox{.}}{2017}]%
        {DBLP:conf/usenix/HahnelCP17}
\bibfield{author}{\bibinfo{person}{Marcus H{\"{a}}hnel},
  \bibinfo{person}{Weidong Cui}, {and} \bibinfo{person}{Marcus Peinado}.}
  \bibinfo{year}{2017}\natexlab{}.
\newblock \showarticletitle{High-Resolution Side Channels for Untrusted
  Operating Systems}. In \bibinfo{booktitle}{\emph{2017 {USENIX} Annual
  Technical Conference, {USENIX} {ATC} 2017, Santa Clara, CA, USA, July 12-14,
  2017.}} \bibinfo{publisher}{{USENIX} Association}, \bibinfo{pages}{299--312}.
\newblock
\urldef\tempurl%
\url{https://www.usenix.org/conference/atc17/technical-sessions/presentation/hahnel}
\showURL{%
\tempurl}


\bibitem[\protect\citeauthoryear{Kirda and Ristenpart}{Kirda and
  Ristenpart}{2017}]%
        {DBLP:conf/uss/2017}
\bibfield{editor}{\bibinfo{person}{Engin Kirda} {and} \bibinfo{person}{Thomas
  Ristenpart}} (Eds.). \bibinfo{year}{2017}\natexlab{}.
\newblock \bibinfo{booktitle}{\emph{26th {USENIX} Security Symposium, {USENIX}
  Security 2017, Vancouver, BC, Canada, August 16-18, 2017}}.
  \bibinfo{publisher}{{USENIX} Association}.
\newblock
\urldef\tempurl%
\url{https://www.usenix.org/conference/usenixsecurity17}
\showURL{%
\tempurl}


\bibitem[\protect\citeauthoryear{Lee, Shih, Gera, Kim, Kim, and Peinado}{Lee
  et~al\mbox{.}}{2017}]%
        {DBLP:conf/uss/0001SGKKP17}
\bibfield{author}{\bibinfo{person}{Sangho Lee}, \bibinfo{person}{Ming{-}Wei
  Shih}, \bibinfo{person}{Prasun Gera}, \bibinfo{person}{Taesoo Kim},
  \bibinfo{person}{Hyesoon Kim}, {and} \bibinfo{person}{Marcus Peinado}.}
  \bibinfo{year}{2017}\natexlab{}.
\newblock \showarticletitle{Inferring Fine-grained Control Flow Inside {SGX}
  Enclaves with Branch Shadowing}, See \citeN{DBLP:conf/uss/2017},
  \bibinfo{pages}{557--574}.
\newblock
\urldef\tempurl%
\url{https://www.usenix.org/conference/usenixsecurity17/technical-sessions/presentation/lee-sangho}
\showURL{%
\tempurl}


\bibitem[\protect\citeauthoryear{Liu, Wang, Nayak, Huang, and Shi}{Liu
  et~al\mbox{.}}{2015}]%
        {DBLP:conf/sp/LiuWNHS15}
\bibfield{author}{\bibinfo{person}{Chang Liu}, \bibinfo{person}{Xiao~Shaun
  Wang}, \bibinfo{person}{Kartik Nayak}, \bibinfo{person}{Yan Huang}, {and}
  \bibinfo{person}{Elaine Shi}.} \bibinfo{year}{2015}\natexlab{}.
\newblock \showarticletitle{ObliVM: {A} Programming Framework for Secure
  Computation}, See \citeN{DBLP:conf/sp/2015}, \bibinfo{pages}{359--376}.
\newblock
\showISBNx{978-1-4673-6949-7}
\urldef\tempurl%
\url{https://doi.org/10.1109/SP.2015.29}
\showDOI{\tempurl}


\bibitem[\protect\citeauthoryear{Ohrimenko, Costa, Fournet, Gkantsidis,
  Kohlweiss, and Sharma}{Ohrimenko et~al\mbox{.}}{2015}]%
        {DBLP:conf/ccs/OhrimenkoCFGKS15}
\bibfield{author}{\bibinfo{person}{Olga Ohrimenko}, \bibinfo{person}{Manuel
  Costa}, \bibinfo{person}{C{\'{e}}dric Fournet}, \bibinfo{person}{Christos
  Gkantsidis}, \bibinfo{person}{Markulf Kohlweiss}, {and}
  \bibinfo{person}{Divya Sharma}.} \bibinfo{year}{2015}\natexlab{}.
\newblock \showarticletitle{Observing and Preventing Leakage in MapReduce}. In
  \bibinfo{booktitle}{\emph{Proceedings of the 22nd {ACM} {SIGSAC} Conference
  on Computer and Communications Security, Denver, CO, USA, October 12-6,
  2015}}, \bibfield{editor}{\bibinfo{person}{Indrajit Ray},
  \bibinfo{person}{Ninghui Li}, {and} \bibinfo{person}{Christopher Kruegel}}
  (Eds.). \bibinfo{publisher}{{ACM}}, \bibinfo{pages}{1570--1581}.
\newblock
\showISBNx{978-1-4503-3832-5}
\urldef\tempurl%
\url{https://doi.org/10.1145/2810103.2813695}
\showDOI{\tempurl}


\bibitem[\protect\citeauthoryear{Ohrimenko, Goodrich, Tamassia, and
  Upfal}{Ohrimenko et~al\mbox{.}}{2014}]%
        {DBLP:conf/icalp/OhrimenkoGTU14}
\bibfield{author}{\bibinfo{person}{Olga Ohrimenko}, \bibinfo{person}{Michael~T.
  Goodrich}, \bibinfo{person}{Roberto Tamassia}, {and} \bibinfo{person}{Eli
  Upfal}.} \bibinfo{year}{2014}\natexlab{}.
\newblock \showarticletitle{The Melbourne Shuffle: Improving Oblivious Storage
  in the Cloud}. In \bibinfo{booktitle}{\emph{Automata, Languages, and
  Programming - 41st International Colloquium, {ICALP} 2014, Copenhagen,
  Denmark, July 8-11, 2014, Proceedings, Part {II}}}.
  \bibinfo{pages}{556--567}.
\newblock
\urldef\tempurl%
\url{https://doi.org/10.1007/978-3-662-43951-7_47}
\showDOI{\tempurl}


\bibitem[\protect\citeauthoryear{Ohrimenko, Schuster, Fournet, Mehta, Nowozin,
  Vaswani, and Costa}{Ohrimenko et~al\mbox{.}}{2016}]%
        {DBLP:conf/uss/OhrimenkoSFMNVC16}
\bibfield{author}{\bibinfo{person}{Olga Ohrimenko}, \bibinfo{person}{Felix
  Schuster}, \bibinfo{person}{C{\'{e}}dric Fournet}, \bibinfo{person}{Aastha
  Mehta}, \bibinfo{person}{Sebastian Nowozin}, \bibinfo{person}{Kapil Vaswani},
  {and} \bibinfo{person}{Manuel Costa}.} \bibinfo{year}{2016}\natexlab{}.
\newblock \showarticletitle{Oblivious Multi-Party Machine Learning on Trusted
  Processors}. In \bibinfo{booktitle}{\emph{25th {USENIX} Security Symposium,
  {USENIX} Security 16, Austin, TX, USA, August 10-12, 2016.}},
  \bibfield{editor}{\bibinfo{person}{Thorsten Holz} {and}
  \bibinfo{person}{Stefan Savage}} (Eds.). \bibinfo{publisher}{{USENIX}
  Association}, \bibinfo{pages}{619--636}.
\newblock
\urldef\tempurl%
\url{https://www.usenix.org/conference/usenixsecurity16/technical-sessions/presentation/ohrimenko}
\showURL{%
\tempurl}


\bibitem[\protect\citeauthoryear{Oleksenko, Trach, Krahn, Silberstein, and
  Fetzer}{Oleksenko et~al\mbox{.}}{2018}]%
        {DBLP:conf/usenix/OleksenkoTKSF18}
\bibfield{author}{\bibinfo{person}{Oleksii Oleksenko}, \bibinfo{person}{Bohdan
  Trach}, \bibinfo{person}{Robert Krahn}, \bibinfo{person}{Mark Silberstein},
  {and} \bibinfo{person}{Christof Fetzer}.} \bibinfo{year}{2018}\natexlab{}.
\newblock \showarticletitle{Varys: Protecting {SGX} Enclaves from Practical
  Side-Channel Attacks}. In \bibinfo{booktitle}{\emph{{USENIX} Annual Technical
  Conference}}. \bibinfo{publisher}{{USENIX} Association},
  \bibinfo{pages}{227--240}.
\newblock


\bibitem[\protect\citeauthoryear{Sasy, Gorbunov, and Fletcher}{Sasy
  et~al\mbox{.}}{2017}]%
        {DBLP:journals/iacr/SasyGF17}
\bibfield{author}{\bibinfo{person}{Sajin Sasy}, \bibinfo{person}{Sergey
  Gorbunov}, {and} \bibinfo{person}{Christopher~W. Fletcher}.}
  \bibinfo{year}{2017}\natexlab{}.
\newblock \showarticletitle{ZeroTrace : Oblivious Memory Primitives from Intel
  {SGX}}.
\newblock \bibinfo{journal}{\emph{{IACR} Cryptology ePrint Archive}}
  \bibinfo{volume}{2017} (\bibinfo{year}{2017}), \bibinfo{pages}{549}.
\newblock
\urldef\tempurl%
\url{http://eprint.iacr.org/2017/549}
\showURL{%
\tempurl}


\bibitem[\protect\citeauthoryear{Shih, Lee, Kim, and Peinado}{Shih
  et~al\mbox{.}}{[n.d.]}]%
        {shih2017t}
\bibfield{author}{\bibinfo{person}{Ming-Wei Shih}, \bibinfo{person}{Sangho
  Lee}, \bibinfo{person}{Taesoo Kim}, {and} \bibinfo{person}{Marcus Peinado}.}
  \bibinfo{year}{[n.d.]}\natexlab{}.
\newblock \showarticletitle{T-SGX: Eradicating controlled-channel attacks
  against enclave programs}. In \bibinfo{booktitle}{\emph{NDSS Symposium 2017
  in San Diego, California.}}
\newblock


\bibitem[\protect\citeauthoryear{Shinde, Chua, Narayanan, and Saxena}{Shinde
  et~al\mbox{.}}{2016}]%
        {DBLP:conf/ccs/ShindeCNS16}
\bibfield{author}{\bibinfo{person}{Shweta Shinde}, \bibinfo{person}{Zheng~Leong
  Chua}, \bibinfo{person}{Viswesh Narayanan}, {and} \bibinfo{person}{Prateek
  Saxena}.} \bibinfo{year}{2016}\natexlab{}.
\newblock \showarticletitle{Preventing Page Faults from Telling Your Secrets}.
  In \bibinfo{booktitle}{\emph{Proceedings of the 11th {ACM} on Asia Conference
  on Computer and Communications Security, AsiaCCS 2016, Xi'an, China, May 30 -
  June 3, 2016}}, \bibfield{editor}{\bibinfo{person}{Xiaofeng Chen},
  \bibinfo{person}{XiaoFeng Wang}, {and} \bibinfo{person}{Xinyi Huang}} (Eds.).
  \bibinfo{publisher}{{ACM}}, \bibinfo{pages}{317--328}.
\newblock
\showISBNx{978-1-4503-4233-9}
\urldef\tempurl%
\url{https://doi.org/10.1145/2897845.2897885}
\showDOI{\tempurl}


\bibitem[\protect\citeauthoryear{Stefanov, van Dijk, Shi, Fletcher, Ren, Yu,
  and Devadas}{Stefanov et~al\mbox{.}}{2013}]%
        {DBLP:conf/ccs/StefanovDSFRYD13}
\bibfield{author}{\bibinfo{person}{Emil Stefanov}, \bibinfo{person}{Marten van
  Dijk}, \bibinfo{person}{Elaine Shi}, \bibinfo{person}{Christopher~W.
  Fletcher}, \bibinfo{person}{Ling Ren}, \bibinfo{person}{Xiangyao Yu}, {and}
  \bibinfo{person}{Srinivas Devadas}.} \bibinfo{year}{2013}\natexlab{}.
\newblock \showarticletitle{Path {ORAM:} an extremely simple oblivious {RAM}
  protocol}. In \bibinfo{booktitle}{\emph{2013 {ACM} {SIGSAC} Conference on
  Computer and Communications Security, CCS'13, Berlin, Germany, November 4-8,
  2013}}. \bibinfo{pages}{299--310}.
\newblock
\urldef\tempurl%
\url{https://doi.org/10.1145/2508859.2516660}
\showDOI{\tempurl}


\bibitem[\protect\citeauthoryear{{Van Bulck}, Weichbrodt, Kapitza, Piessens,
  and Strackx}{{Van Bulck} et~al\mbox{.}}{2017}]%
        {DBLP:conf/uss/BulckWKPS17}
\bibfield{author}{\bibinfo{person}{Jo {Van Bulck}}, \bibinfo{person}{Nico
  Weichbrodt}, \bibinfo{person}{R{\"{u}}diger Kapitza}, \bibinfo{person}{Frank
  Piessens}, {and} \bibinfo{person}{Raoul Strackx}.}
  \bibinfo{year}{2017}\natexlab{}.
\newblock \showarticletitle{Telling Your Secrets without Page Faults: Stealthy
  Page Table-Based Attacks on Enclaved Execution}, See
  \citeN{DBLP:conf/uss/2017}, \bibinfo{pages}{1041--1056}.
\newblock
\urldef\tempurl%
\url{https://www.usenix.org/conference/usenixsecurity17/technical-sessions/presentation/van-bulck}
\showURL{%
\tempurl}


\bibitem[\protect\citeauthoryear{Wang, Chen, Pan, Zhang, Wang, Bindschaedler,
  Tang, and Gunter}{Wang et~al\mbox{.}}{2017}]%
        {DBLP:conf/ccs/WangCPZWBTG17}
\bibfield{author}{\bibinfo{person}{Wenhao Wang}, \bibinfo{person}{Guoxing
  Chen}, \bibinfo{person}{Xiaorui Pan}, \bibinfo{person}{Yinqian Zhang},
  \bibinfo{person}{XiaoFeng Wang}, \bibinfo{person}{Vincent Bindschaedler},
  \bibinfo{person}{Haixu Tang}, {and} \bibinfo{person}{Carl~A. Gunter}.}
  \bibinfo{year}{2017}\natexlab{}.
\newblock \showarticletitle{Leaky Cauldron on the Dark Land: Understanding
  Memory Side-Channel Hazards in {SGX}}. In
  \bibinfo{booktitle}{\emph{Proceedings of the 2017 {ACM} {SIGSAC} Conference
  on Computer and Communications Security, {CCS} 2017, Dallas, TX, USA, October
  30 - November 03, 2017}}, \bibfield{editor}{\bibinfo{person}{Bhavani~M.
  Thuraisingham}, \bibinfo{person}{David Evans}, \bibinfo{person}{Tal Malkin},
  {and} \bibinfo{person}{Dongyan Xu}} (Eds.). \bibinfo{publisher}{{ACM}},
  \bibinfo{pages}{2421--2434}.
\newblock
\showISBNx{978-1-4503-4946-8}
\urldef\tempurl%
\url{https://doi.org/10.1145/3133956.3134038}
\showDOI{\tempurl}


\bibitem[\protect\citeauthoryear{Williams and Sion}{Williams and Sion}{2008}]%
        {DBLP:conf/ndss/WilliamsS08}
\bibfield{author}{\bibinfo{person}{Peter Williams} {and} \bibinfo{person}{Radu
  Sion}.} \bibinfo{year}{2008}\natexlab{}.
\newblock \showarticletitle{Usable {PIR}}. In
  \bibinfo{booktitle}{\emph{Proceedings of the Network and Distributed System
  Security Symposium, {NDSS} 2008, San Diego, California, USA, 10th February -
  13th February 2008}}. \bibinfo{publisher}{The Internet Society}.
\newblock
\urldef\tempurl%
\url{http://www.isoc.org/isoc/conferences/ndss/08/papers/09_usable_pir.pdf}
\showURL{%
\tempurl}


\bibitem[\protect\citeauthoryear{Xu, Cui, and Peinado}{Xu
  et~al\mbox{.}}{2015}]%
        {DBLP:conf/sp/XuCP15}
\bibfield{author}{\bibinfo{person}{Yuanzhong Xu}, \bibinfo{person}{Weidong
  Cui}, {and} \bibinfo{person}{Marcus Peinado}.}
  \bibinfo{year}{2015}\natexlab{}.
\newblock \showarticletitle{Controlled-Channel Attacks: Deterministic Side
  Channels for Untrusted Operating Systems}, See \citeN{DBLP:conf/sp/2015},
  \bibinfo{pages}{640--656}.
\newblock
\showISBNx{978-1-4673-6949-7}
\urldef\tempurl%
\url{https://doi.org/10.1109/SP.2015.45}
\showDOI{\tempurl}


\bibitem[\protect\citeauthoryear{Zheng, Dave, Beekman, Popa, Gonzalez, and
  Stoica}{Zheng et~al\mbox{.}}{2017}]%
        {DBLP:conf/nsdi/ZhengDBPGS17}
\bibfield{author}{\bibinfo{person}{Wenting Zheng}, \bibinfo{person}{Ankur
  Dave}, \bibinfo{person}{Jethro~G. Beekman}, \bibinfo{person}{Raluca~Ada
  Popa}, \bibinfo{person}{Joseph~E. Gonzalez}, {and} \bibinfo{person}{Ion
  Stoica}.} \bibinfo{year}{2017}\natexlab{}.
\newblock \showarticletitle{Opaque: An Oblivious and Encrypted Distributed
  Analytics Platform}. In \bibinfo{booktitle}{\emph{14th {USENIX} Symposium on
  Networked Systems Design and Implementation, {NSDI} 2017, Boston, MA, USA,
  March 27-29, 2017}}. \bibinfo{pages}{283--298}.
\newblock
\urldef\tempurl%
\url{https://www.usenix.org/conference/nsdi17/technical-sessions/presentation/zheng}
\showURL{%
\tempurl}


\end{thebibliography}


\begin{thebibliography}{1}

\bibitem{bowman:reasoning}
M.~Bowman, S.~K. Debray, and L.~L. Peterson.
\newblock Reasoning about naming systems.
\newblock {\em ACM Trans. Program. Lang. Syst.}, 15(5):795--825, November 1993.

\bibitem{braams:babel}
J.~Braams.
\newblock Babel, a multilingual style-option system for use with latex's
  standard document styles.
\newblock {\em TUGboat}, 12(2):291--301, June 1991.

\bibitem{clark:pct}
M.~Clark.
\newblock Post congress tristesse.
\newblock In {\em TeX90 Conference Proceedings}, pages 84--89. TeX Users Group,
  March 1991.

\bibitem{herlihy:methodology}
M.~Herlihy.
\newblock A methodology for implementing highly concurrent data objects.
\newblock {\em ACM Trans. Program. Lang. Syst.}, 15(5):745--770, November 1993.

\bibitem{Lamport:LaTeX}
L.~Lamport.
\newblock {\em LaTeX User's Guide and Document Reference Manual}.
\newblock Addison-Wesley Publishing Company, Reading, Massachusetts, 1986.

\bibitem{salas:calculus}
S.~Salas and E.~Hille.
\newblock {\em Calculus: One and Several Variable}.
\newblock John Wiley and Sons, New York, 1978.

\end{thebibliography}

\end{document}